\documentclass[
reprint,
superscriptaddress,
nofootinbib,
 amsmath,amssymb,
 aps,
prd,
]{revtex4-2}

\usepackage{tikz}
\usepackage{tikz-3dplot}
\usetikzlibrary{arrows.meta,positioning}
\usepackage[hidelinks,colorlinks=true,linkcolor=blue,citecolor=blue]{hyperref}
\newcommand\aap{A\&A}                

\newcommand\apjl{ApJ}                
\newcommand\apjs{ApJS}               

\usetikzlibrary{fit}
\usepackage[usenames, dvipsnames]{xcolor}

\usepackage[normalem]{ulem}
\usepackage{natbib}
\usepackage{CJK}
\usepackage{graphicx}
\usepackage{dcolumn}
\usepackage{bm}

\usepackage{aas_macros}
\usepackage{orcidlink}
\usepackage{tikz}
\usepackage{tikz-3dplot}
\usetikzlibrary{arrows.meta, decorations.markings, calc}

\begin{document}


\title{Dynamics of Relativistic Binaries in Structured and Stochastic Environments:\\ A Lagrange--Fourier--Hansen Framework.}

\author{Lorenz Zwick}
\email{lorenz.zwick@nbi.ku.dk}
\author{Conor Dyson}
\author{Brian C. Seymour} 
\affiliation{Center of Gravity, Niels Bohr Institute, Blegdamsvej 17, 2100 Copenhagen, Denmark.}
\author{János Takátsy}
\affiliation{Institut für Physik und Astronomie, Universität Potsdam, Haus 28, Karl-Liebknecht-Str. 24-25, Potsdam, Germany}
\author{Johan Samsing}
\affiliation{Center of Gravity, Niels Bohr Institute, Blegdamsvej 17, 2100 Copenhagen, Denmark.}


%


\date{\today}

\begin{abstract}
We develop a general framework to characterize non-vacuum perturbations to relativistic binaries in the gravitational-wave (GW) driven regime, for use in GW parameter estimation studies. The effect of smooth, structured and stochastic perturbations to the binary's motion is reduced to a resonant spectral projection defined on a rolling averaging window, with weights given by Hansen coefficients. This is combined with practical criteria for identifying and evaluating the corresponding dynamical response to perturbations, starting from either analytical models or numerical simulations of binaries in environments. The result is a set of coupled ODEs for the orbital elements that capture epi-cyclic, apsidal and nodal resonances, consistently incorporate feedback from radiation reaction and can be solved efficiently on a coarse time grid. We demonstrate the practical application of the framework in two representative astrophysical scenarios: a compact binary in a variable tidal field and an extreme-mass-ratio inspiral in an accretion disk. We propose the Lagrange-Fourier-Hansen framework as a unified tool for modeling environmental effects in GW templates for eccentric and precessing binary sources, and particularly for bridging the gap between phenomenological prescriptions and realistic models of binaries in environments.
\end{abstract}

\maketitle


\section{Introduction}
\label{sec:introduction}
Unlike electromagnetic radiation which is scattered, absorbed, and distorted by intervening matter, gravitational waves (GW) propagate through cosmological distances virtually unimpeded. As a consequence, the GW phase observed at Earth remains a direct map of the binary source's orbital dynamics, down to the precise location of the two component masses. This direct correspondence between GW phase and binary dynamics is precisely why such sustained effort is required to develop high-accuracy schemes for solving the field equations of General Relativity (GR) across the complementary regimes of post-Newtonian (PN) theory \cite{1985damour,1991damour,1998jaranowski,1999jaranowski,1999buonanno,2001damoureob,2010barack,2012akcay,2014pan,2020ossokine}, self-force methods  \cite{1997Quinn, 1997mino, 2009Pound, 2017vandeMeentGen, 2021fyyWardell}, and numerical relativity \cite{2005pretor,ajith,purrer,pfeffer}. The resulting waveform templates underpin the matched-filtering methods that form the backbone of GW data analysis and have enabled the extraordinary advances of GW astronomy in the LIGO/Virgo/KAGRA era \cite{LIGOdetector,abbottastrophysics,abbottmultimessenger,abbottGW151226,abbottGW190412,Abbot2023}.
However, the development of these methods has largely focused on the case of GW sources evolving in vacuum, an assumption which is known to break down for astrophysical sources.

Indeed, from stellar-mass black hole (BH) binaries \citep{2002belczynski,2007oleary,2008Sadowski,2016antonini,2017vitale,kavanagh2020,zevin2020,2021zevin,2021kimball,2023santini, portegieszwart2000, Lee:2010in,
2010MNRAS.402..371B, 2013MNRAS.435.1358T, 2014MNRAS.440.2714B,2015PhRvL.115e1101R, 2015ApJ...802L..22R, 2016PhRvD..93h4029R, 2016ApJ...824L...8R,2016ApJ...824L...8R, 2017MNRAS.464L..36A, 2017MNRAS.469.4665P, Samsing18, 2018MNRAS.tmp.2223S, 2020PhRvD.101l3010S, 2021MNRAS.504..910T, 2013ApJ...773..187N, 2014ApJ...785..116L, 2016ApJ...816...65A, 2016MNRAS.456.4219A, 2017ApJ...836...39S, 2018ApJ...864..134R, 2019ApJ...883...23H, 2021MNRAS.502.2049L,2022MNRAS.511.1362T,2017ApJ...835..165B,  2017MNRAS.464..946S, 2017arXiv170207818M, 2020ApJ...898...25T, 2022Natur.603..237S,Trani:2023oqa, Fabj24} to massive BH binaries and extreme-mass-ratio inspirals (EMRIs) \cite{amaro-seoane2007,2017science,2017lisa,2019lisa,2022lisaastro,2024lisa}, the assembly of GW sources typically requires astrophysical processes that occur in dense and dynamically rich environments, including galactic nuclei, dark matter cusps, stellar clusters, and various forms of accretion discs. Crucially, these environments can remain dynamically coupled to the binary throughout the observable GW inspiral, giving rise to {\it environmental effects} (EEs), departures from vacuum binary evolution induced by gas, stars, external gravitational potentials or other ambient structure, that are reflected in the emitted GWs. Over the last three decades, a broad literature has developed around EEs in GW sources \cite{1993chakrabarti,1995ryan,2008barausse,2007levin,kocsis,2014barausse,inayoshi2017,2017meiron,2017Bonetti,2019alejandro,2019randall,Yu:2020dlm,2020cardoso,DOrazioGWLens:2020,Yu:2021dqx,2022liu,2022xuan,garg2022,2022cole,2022chandramouli,2022sberna,2023zwick,2023Tiede,2024dyson,2022destounis,2022cardoso,2020caputo,2023arXiv231016799L,2024zwicknovel,2021andrea,2024basu,2024santoro,2025dyson}. Several seminal studies show that gas drag, accretion, dynamical friction, stellar scattering, and external tidal perturbations can, in selected regimes, produce secular modifications to the binary inspiral and hence to the observed waveform. The scientific case for LISA \cite{2024lisa} and third-generation ground-based detectors \cite{2020maggioreet,2023arXiv230613745E} has significantly accelerated this program, as long observation times and increased sensitivity result in weaker departures from vacuum evolution becoming observable. Up to recent years, the main methodology to treat EEs has been through phenomenological models that induce power--law corrections to the GW phase of the form $\delta \psi_{\rm GW} \sim f_{\rm GW}^{n}$, where $f_{\rm GW}^{n}$ is the GW frequency and $n$ is a power law index \cite{2014barausse,2020cardosoee}. This approach has been extremely useful: First, such models are analytically transparent and easily related to common EE or modifications to GR. Second, they are computationally inexpensive and straightforward to incorporate into Fourier domain waveform templates. Therefore, simple dephasing prescriptions have provided the first estimates for assessing whether a given EE might be observable at all.

However, more recent studies have started to highlight how this methodology is fundamentally limited in its ability to realistically model the coupling between relativistic binaries and their astrophysical environments, as well as any non-vacuum effect that does not reduce to a smooth energy flux. The reasons are two-fold: firstly, the dynamics of relativistic binaries is itself complex, and effects like eccentricity and precession influence GW emission in ways that go beyond simple parametrization \cite{2009yunes,2022knee,2025gamboa}. Secondly, realistic environmental perturbations are most often characterized by structured, variable, or even stochastic forcings that do not reduce to a simple description. Of particular note are perturbations arising in gas embedded binaries. In realistic scenarios, these are characterized by time-varying or stochastic perturbations with broad--band frequency spectra rather than the commonly used constant torque or dynamical friction \cite{2005nelson,Roedig_Trqs+2012,2022zwick,ONeill2024,2024zwick,2025Zwick_ecc,2025copparoni,2026arXiv260420971G}. Both of these limitations are especially acute for {\it eccentric and precessing binaries}, whose dynamics are built from a hierarchy of orbital harmonics \cite{peters1964,2009yunes,2018klein}. For such sources, there is no reason to expect that the leading order imprint on the GW is a single secular dephasing prescription \cite{2025Zwick_ecc}. Instead, it is in principle necessary to treat the full spectrum of perturbations and its projection onto the binary's orbital and precessional modes. This is particularly important because eccentricity is a natural outcome in several source classes where strong environmental interactions are expected, including dynamically assembled stellar-mass binaries, extreme-mass-ratio inspirals, and massive black hole binaries in gaseous or stellar environments. Consequently, \textit{phenomenological treatments of EEs risk missing the most informative signatures precisely in the systems where those signatures are most likely to arise}. 

If EEs are to be modeled accurately, what is required is a framework that can treat {\it generic perturbing forces}, including forces extracted directly from numerical simulations, and translate them into controlled corrections to GW observables for eccentric and precessing binaries. The aim of this work is to develop such a framework and provide a general and computationally efficient tool to characterize the effect of perturbative forces on waveform observables, from analytic models to simulation results. To achieve this aim, we will combine several historical methods of celestial mechanics and apply them to relativistic orbits. Firstly, the Lagrange Planetary Equations \cite{euler,gauss,2014poisson} (LPE) to treat orbital perturbations. Secondly, aspects of Fourier spectral analysis \cite{fourier} to treat structured, broadband and stochastic forces. Thirdly, Hansen coefficients \cite{hansen}, as a useful basis to project orbital perturbations onto epicycles of the mean anomaly.  Therefore, we refer to the results of this work as the Lagrange-Fourier-Hansen (LFH) framework. To facilitate practical applications, we are currently developing an implementation of the framework as a Python package, \texttt{AstroWaveforms}. The package is designed to operate as a black-box interface, taking as input a binary vacuum trajectory and a perturbing force and returning the corresponding orbital evolution and waveform observables. The release of the \texttt{AstroWaveforms} package, including benchmarking, additional use cases, and inference studies, will be presented in a forthcoming publication.

The remainder of this paper is organized as follows: In section \ref{sec:framework_overview}, we introduce the main ideas behind the framework, outline its assumptions and present an overview of the full derivation. In section \ref{sec:practice}, we turn to the practical implementation of the framework, including the spectral decomposition of force time series, mode truncation, identification of stationary points, and integration of the resulting orbital evolution equations. We illustrate the steps using worked examples involving a time-dependent tidal perturbation and a fluctuating accretion-disc torque. In section \ref{sec:discussion}, we discuss the applications and important extensions of the framework. Finally, we present some concluding remarks in section \ref{sec:conclusion}. Technical derivations, including the complete resonant response formalism and tables of Hansen coefficients, are collected in the Appendices \ref{Appendix:Resonant_response} and \ref{Appendix:Hansen_coeff}.
\begin{figure}
    \centering
    \includegraphics[width=0.75\linewidth]{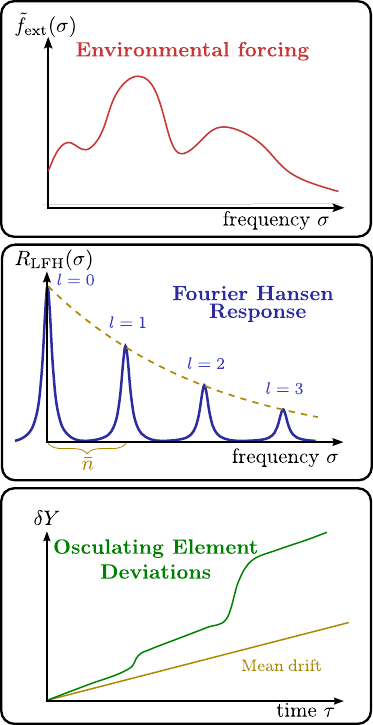}
    \caption{Illustration of the basic elements underlying the LFH framework. An environmental perturbation is treated in Fourier space, and in general can have support over a broad range of frequencies. The secular dynamics of the binary only couples with select resonant response channels, with weights given by Hansen coefficients. The generic result is a rich dynamical evolution, characterized by drifts, resonant jumps and feedback with radiation reaction. A framework to efficiently model this rich dynamics can be used to enrich gravitational waveform templates with smoking gun signatures of binary environments. }
    \label{fig:LFH-illustration}
\end{figure}
\section{The LFH framework in theory}
\label{sec:framework_overview}
\subsection{Goal and basic insight}

The aim of the LFH framework is to provide a general and computationally efficient map between a perturbing force, including forces extracted from numerical simulations, and the orbital dynamics of an eccentric, precessing binary in the GW-dominated regime. In particular, we want to model the binary phase with sufficient precision, since it is directly tied to GW observables. While such a map can in principle be obtained by direct numerical integration, this requires resolving timescales much shorter than the orbital period, making it impractical for GW data analysis and parameter inference. Instead, as is often the case in celestial mechanics, we want to isolate the long-term response of the system while averaging over non-secular orbital oscillations. The starting insight is that eccentric and precessing binaries are intrinsically multi-frequency systems, with dynamics built from a hierarchy of modes:
\begin{align}
\text{Orbital mode} \sim l\,\bar{n} + k_{\omega}\,\dot{\omega} + k_{\Omega}\,\dot{\Omega} + k_{\iota}\, \dot \iota .
\end{align}
Here, $\bar{n}$ is the mean motion, while $\dot{\omega}$, $\dot{\Omega}$ and $\dot \iota$ are the apsidal, nodal and inclination precession rates. The index $l$ labels orbital harmonics arising from eccentric motion, while the $k_i$ denote modes associated with precession of the orbital plane, e.g.~due to spin-orbit and spin-spin couplings \cite{blanchet2014,2017will,2023gerosa,2023fumagalli}. The presence of several frequencies immediately implies that external perturbations can, in principle, cause a dynamical response across any of these orbital modes. Indeed, the secular evolution is controlled by components of the forcing that satisfy resonance conditions of the form:
\begin{align}
    \sigma+l\,\bar{n} + k_{\omega}\,\dot{\omega} + k_{\Omega}\,\dot{\Omega} + k_{\iota}\, \dot \iota  \simeq 0,
\end{align}
where the spectrum of the external perturbation has support near the frequency $\sigma$. Thus, the goal of this work is to identify and evaluate these resonant channels systematically for binaries that are already driven by radiation reaction and exhibit relativistic precession. An illustration of the goals and basic insight of the framework is given in Fig.~\ref{fig:LFH-illustration}.
\begin{figure}
\tdplotsetmaincoords{70}{110}
\begin{tikzpicture}[tdplot_main_coords,scale=4.4]
  \pgfmathsetmacro{\r}{.8}
  \pgfmathsetmacro{\O}{45}
  \pgfmathsetmacro{\i}{30}
  \pgfmathsetmacro{\f}{35}
  \coordinate (O) at (0,0,0);

  \draw [->, very thick] (O) -- (2,0,0) node[anchor=north east] {$x$};
  \draw [->, very thick] (O) -- (0,1,0) node[anchor=north west] {$y$};
  \draw [->, very thick] (O) -- (0,0,1) node[anchor=south] {$z$};

  \tdplotdrawarc[dashed]{(O)}{\r}{0}{360}{}{}
  \tdplotsetrotatedcoords{\O}{0}{0}

  \draw [tdplot_rotated_coords] (-1,0,0) -- (1,0,0);
  \tdplotdrawarc[->]{(O)}{.33*\r}{0}{\O}{anchor=north}{$\Omega$}
  \tdplotsetrotatedcoords{-\O}{\i}{0}
  \tdplotdrawarc[tdplot_rotated_coords]{(O)}{\r}{0}{360}{}{}  

  \begin{scope}[tdplot_rotated_coords]
    \draw[->] (O) -- (0,0,1) node [above] {};
    \draw (1,0,0) -- (-1,0,0);
    \tdplotdrawarc[->]{(O)}{.33*\r}{90}{180}{anchor=west}{$\omega$}
    \coordinate (P) at (180+\f:\r);
    \draw (O) -- (P);
    \tdplotdrawarc[->]{(O)}{.33*\r}{180}{180+\f}{anchor=south west}{$\nu$}
  \end{scope}

  \tdplotsetrotatedcoords{-\O+\f}{\i}{0}
  \tdplotsetrotatedcoordsorigin{(P)}

  \begin{scope}[tdplot_rotated_coords,scale=.2,thick]
    \draw [->] (P) -- (-1,0,0) node [right] {$\hat{\mathbf{r}}$};
    \draw [->] (P) -- (0,-1,0) node [above] {$\hat{\mathbf{s}}$};
    \draw [->] (P) -- (0,0,1) node [above] {$\hat{\mathbf{w}}$};
    \fill (P) circle (.33ex);
  \end{scope}

  \tdplotsetthetaplanecoords{-\f}
  \tdplotdrawarc[tdplot_rotated_coords,->]{(O)}{.75*\r}{0}{\i}{anchor=south}{$\iota$}
\end{tikzpicture}
\caption{Orbital geometry of an eccentric and inclined binary. The orbit is defined by the longitude of the ascending node $\Omega$, inclination $\iota$, argument of pericenter $\omega$ and true anomaly $\nu$. The elements are understood as osculating, where the perturbative force is decomposed along the radial, tangential and normal vectors $(\hat {\mathbf r},\hat {\mathbf s},\hat {\mathbf w})$. Note that this basis is attached to the orbit and is therefore time dependent.}
\label{fig:orbit_sketch_main}
\end{figure}
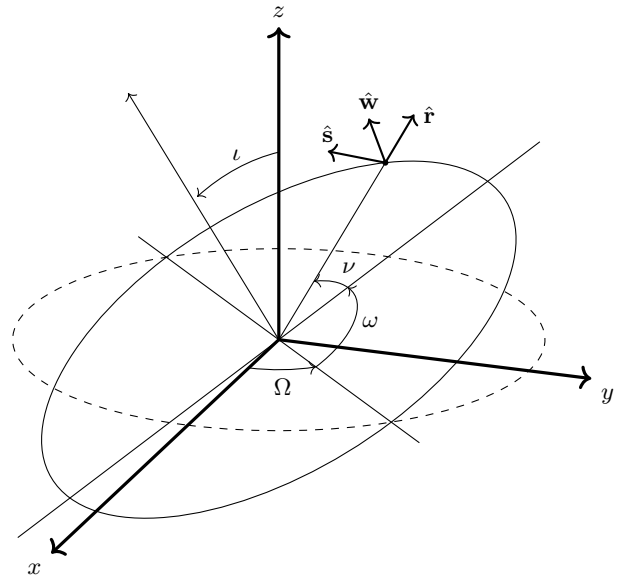

\subsection{Assumptions}

In order to describe the response of a binary to environmental perturbations, we first require a concrete representation of the vacuum orbital dynamics. For the derivation presented in this work we, assume that the vacuum orbit can be approximated as a rigidly precessing Keplerian conic section with orbital elements driven by orbit averaged radiation reaction and conservative precession. We parameterize the instantaneous orbit in terms of the classic osculating elements:
\begin{align}
\mathbf{Y} = (p, e, \omega, \Omega, \iota,M).
\end{align}
where $p$ is the semi-latus rectum, $e$ the eccentricity, $\omega$ the argument of pericenter, $\Omega$ the longitude of the ascending node, $\iota$ the inclination and $M$ is the mean anomaly (see Fig.~\ref{fig:orbit_sketch_main}), with precession rates $\dot \omega$, $\dot \Omega$ and $\dot \iota$. The radius of the orbit takes the conic section form:
\begin{align}
    r=\frac{p}{1+e \cos \nu}\,,
\end{align}
where the relation between true anomaly $\nu$ and mean anomaly $M$ is given by Kepler's equation.
We will discuss an extension to the Quasi-Keplerian or self-force theory set of elements in section \ref{sec:discussion}. For now, we simply note that a precessing conic-section parametrization captures a full set of relativistic orbital frequencies, i.e. radial, apsidal and nodal (see discussion in section \ref{sec:implications}). Additionally, using classic orbital elements makes the connections with numerical simulations of binaries in astrophysical environments more transparent, as the latter field typically works under Newtonian assumptions.

To describe the impact of perturbations, we make use of the LPE, which for our parametrization retain the standard Gauss form (a short proof is given in Appendix \ref{Appendix:Resonant_response}):
\begin{align}
\dot p
&=
2\sqrt{\frac{p^3}{Gm}}
\frac{S}{1+e\cos\nu},
\\[1ex]
\dot e
&=
\sqrt{\frac{p}{Gm}}
\left[
R\sin\nu
+
S\,
\frac{2\cos\nu+e(1+\cos^2\nu)}
     {1+e\cos\nu}
\right],
\\[1ex]
\dot\iota
&=
\sqrt{\frac{p}{Gm}}
\frac{\cos(\omega+\nu)}
     {1+e\cos\nu}
\,W,
\\[1ex]
\dot\Omega
&=
\sqrt{\frac{p}{Gm}}
\frac{\sin(\omega+\nu)}
     {\sin\iota\,(1+e\cos\nu)}
\,W,
\\[1ex]
\dot\omega
&=
\sqrt{\frac{p}{Gm}}
\left[
-\frac{\cos\nu}{e}\,R
+
\frac{\sin\nu}{e}
\frac{2+e\cos\nu}
     {1+e\cos\nu}
\,S
\right]\nonumber
\\ &-
\cos\iota\,\dot\Omega,
\\[1ex]
\dot M
&=
\bar n
-
\frac{1-e^2}{e}
\sqrt{\frac{p}{Gm}}
\bigg[
\left(
\cos\nu-\frac{2e}{1-e^2}
\frac{r}{p}
\right)R \nonumber \\
&-
\sin\nu
\left(
1+\frac{1}{1+e\cos\nu}
\right)S
\bigg],
\end{align}
and can be adapted to relativistic parameterizations of the orbital motion, which we discuss in section \ref{sec:implications}. Here the perturbing is acceleration decomposed into radial ($R$), azimuthal ($S$), and normal components ($W$). Note that, for a rigidly precessing ellipse, only the mean anomaly $M$ has an intrinsic response that does not vanish in the absence of perturbing forces, through the mean motion $\bar n$. Crucially, we make no assumptions about the structure of the perturbing force beyond requiring that it remains weak compared to the vacuum conservative dynamics. This allows us to treat the environmental response within first-order perturbation theory, where the right hand side of the LPE are replaced with functions evaluated at the the vacuum carrier. This assumption will then be relaxed by turning the resulting equations into a set of coupled ODEs.

\subsection{Overview of the derivation}
The thorough derivation underlying the LFH framework is presented in Appendix \ref{Appendix:Resonant_response}. Here, we provide an overview that highlights key ingredients and justifies the methodological choices. References to sections and equations in the Appendix are given for readers interested in the full derivation. Throughout, quantities with a subscript ``${\rm c}$'' denote the vacuum orbital evolution in the absence of perturbations, which we refer to as ``carrier'' orbit. The accelerations $(R_{\rm c},S_{\rm c},W_{\rm c})$ are similarly evaluated on the carrier orbit. The subscript ``${\rm p}$'' denotes the small shifts caused by perturbations. The overall elements evolve as the sum of vacuum and perturbed values.

\subsubsection{Adiabatic window and carrier orbit}
As shown in Appendix \ref{Appendix:Resonant_response:adiabatic}, the first step of the derivation is to isolate the instantaneous response of the binary to environmental forcing from long-term evolution driven by radiation reaction. This is achieved by introducing a local adiabatic time window centered around an epoch $\tau$, within which the orbital motion may be treated as conservative. In practice, we consider intervals of the form: 
\begin{align}
t
\in \mathcal{I}_{\tau}\equiv
\left[
\tau-\frac{\Delta t_W}{2},
\,
\tau+\frac{\Delta t_W}{2}
\right],
\end{align}
with duration satisfying: 
\begin{align}
T_{\rm orb}(\tau)
\ll
\Delta t_W
\ll
T_{\rm rr}(\tau),
\end{align}
with the additional constraint that $\Delta t_W >> 1/\dot \omega$, such that the system can complete many orbital and precession cycles. Here $t$ is an averaging variable valid only within the window, while $\tau$ represents the physical time under which the binary is evolving. Then, $T_{\rm orb}$ is the orbital timescale and $T_{\rm rr}$ is the radiation-reaction timescale at the time $\tau$. The binary therefore completes many orbital cycles within the window while the radiation-reaction driven in-spiral remains negligible. Inside each adiabatic window, the vacuum binary is approximately described by a precessing carrier orbit specified by the orbital elements $\mathbf Y_{\rm c}$ and the corresponding mean motion $\bar n_{\rm c}$ (see again Fig. \ref{fig:orbit_sketch_main}). Within the adiabatic window, the orbital frequencies can be treated as constant. Therefore, the angular variables evolve linearly in $t$:
\begin{align}
\omega_{\rm c}(t)
&\simeq
\omega_{\rm c}(\tau)
+
\dot\omega_{\rm c}(\tau)
(t-\tau),
\\
\Omega_{\rm c}(t)
&\simeq
\Omega_{\rm c}(\tau)
+
\dot\Omega_{\rm c}(\tau)
(t-\tau),
\\
\iota_{\rm c}(t)
&\simeq
\iota_{\rm c}(\tau)
+
\dot\iota_{\rm c}(\tau)
(t-\tau),
\\
M_{\rm c}(t)
&\simeq
M_{\rm c}(\tau)
+
\bar n_{\rm c}(\tau)
(t-\tau).
\end{align}

\subsubsection{Fourier--Hansen decomposition of the forcing}
\label{sec:overview:FH}
As shown in Appendix \ref{Appendix:Resonant_response:Hansen}, the next step is to express both the perturbing force and the orbital response in a form that makes their frequency content explicit. Within the adiabatic window, we introduce a windowed Fourier decomposition of the perturbing accelerations:
\begin{align}
\tilde R_{\rm c}(\tau;\sigma)
&\equiv
\int_{\mathcal I_\tau}
R_{\rm c}(t')\,
e^{-i\sigma t'}
\,dt',
\\
\tilde S_{\rm c}(\tau;\sigma)
&\equiv
\int_{\mathcal I_\tau}
S_{\rm c}(t')\,
e^{-i\sigma t'}
\,dt',
\\
\tilde W_{\rm c}(\tau;\sigma)
&\equiv
\int_{\mathcal I_\tau}
W_{\rm c}(t')\,
e^{-i\sigma t'}
\,dt'.
\label{eq:overview_force_spectrum}
\end{align}
These acceleration amplitudes encode the spectral support of the environmental forcing inside each adiabatic window. Since the coefficients depend explicitly on the window center $\tau$, they also capture any slow evolution of the forcing along the inspiral. In practice, these windowed integrals are performed with a windowing function:
\begin{align}
    \Delta t_{W}\equiv\int_{-\infty}^{\infty}  \mathcal W_{\mathcal I}(t) \,dt ,
\end{align}
and the inverse is performed via the corresponding windowing kernel in frequency space, $\tilde {\mathcal{W}}_{\mathcal I}(\sigma)$, as shown in Eq.~\ref{eq:local_force_decomposition}.

The LPE contain explicit dependencies on the orbital radius $r(\nu_{\rm c})$ together with trigonometric functions of the true anomaly $\nu_{\rm c}$. To expose the harmonic structure of these terms, we expand them using Hansen coefficients \cite{hansen}:
\begin{align}
\left(
\frac{r}{a}
\right)^j
e^{ik\nu}
=
\sum_{l=-\infty}^{\infty}
X_l^{jk}(e)\,
e^{ilM},
\label{eq:overview_hansen_expansion}
\end{align}
with:
\begin{align}
X_l^{jk}(e)
=
\frac{1}{2\pi}
\int_0^{2\pi}
\left(
\frac{r}{a}
\right)^j
e^{ik\nu-ilM}
\,dM.
\end{align}
This decomposition is a crucial step in the derivation. It rewrites the orbital dependence of the LPE as harmonics of the mean anomaly $M_{\rm c}$, whose evolution is locally linear in time.
The Hansen coefficients therefore serve both to make explicit the eccentricity-dependent coupling strengths and the orbital modes through which the environmental forcing couples to the binary motion. Additionally, Hansen coefficients can be pre-tabulated for efficiency. A table of Hansen coefficients is given in Appendix \ref{Appendix:Hansen_coeff}.

Combining the Fourier decomposition with the Hansen expansion, each orbital element evolves as a sum over harmonics weighted by the forcing spectrum. For example, the evolution of the semi-latus rectum becomes:
\begin{align}
\dot p_{\rm p}
&=
2a_{\rm c}
\sqrt{\frac{p_{\rm c}}{Gm}}
\sum_l
X_l^{1,0}
\int
\tilde S_{\rm c}(\tau;\sigma)\,
e^{i\Phi_l^{(0)}(\tau;t)}
\,d\sigma,
\label{eq:overview_pdot_fourier_hansen}
\end{align}
where the full results are given in Eqs.~\ref{eq:lpe_p_fourier} to \ref{eq:lpe_omega_fourier}.
The phases:
\begin{align}
\Phi_l^{(q)}(\tau;t)
\equiv
\sigma t
+
lM_{\rm c}
+
q\omega_{\rm c},
\qquad
q=0,\pm1.
\label{eq:overview_total_phase}
\end{align}
are labeled by the orbital harmonics generated by the Hansen expansion via the index $l$, while the index $q$ labels the apsidal sidebands that appear explicitly in the LPE for  $\Omega$ and $\iota$.  A short summary of the interpretation of these modes is given in Table \ref{tab:mode_interp}.  Note that the force spectral amplitudes are themselves complex quantities, carrying both a magnitude and a phase.
\begin{table}[t]
\centering
\begin{tabular}{lll}
\hline
Mode index & Interpretation & Couples to \\
\hline
$l=0$ & Mean motion & Window average \\
$l=\pm1,\pm2,\ldots$ & Epicycles & Perturbations at $l\bar n$ \\
$q=\pm1$ & Apsidal sidebands  & Perturbations at $q\dot\omega$ \\
\hline
\end{tabular}
\caption{
Interpretation of the mode indices. The index $l$ labels the epicycles generated by expanding the orbital motion in Hansen coefficients. The $l=0$ mode describes the mean motion of the binary that couples to the window-averaged (or ''DC'') component of the forcing, while $l=\pm1,\pm2,\ldots$ describe harmonics of the orbit that couple to forcing components oscillating at $l\bar n$. The index $q$ labels apsidal sidebands, which couple to forcing components oscillating at $q\dot\omega$.
}
\label{tab:mode_interp}
\end{table}

\subsubsection{Resonance selection through window averaging}
At this stage, the LPE contain many oscillatory contributions which do not lead to secular evolution. To identify the terms that lead to secular evolution, and in particular the possible resonances, we average the evolution equations over each adiabatic window. We define the local average over the interval $\mathcal I_\tau$ (Eq.~\ref{eq:local_secular_average}) as:
\begin{align}
\langle
\dot x_{\rm p}
\rangle_{\mathcal I}(\tau)
\equiv
\frac{1}{\Delta t_W}
\int_{\mathcal I_\tau}
\dot x_{\rm p}(\tau;t)\,dt,
\label{eq:overview_local_average}
\end{align}
where $x_{\rm p}$ denotes a generic orbital element. Substituting the Fourier--Hansen decomposition (section \ref{sec:overview:FH}) into the averaging integral gives a double integral, which can be solved analytically. The full derivation is presented in Appendix \ref{Appendix:Resonant_response:selection}. The key insight is that within each adiabatic window, the orbital phases evolve approximately linearly in time (see Eq.~\ref{eq:phase_linearization}). The averaging integral therefore acts as a spectral sinc filter (see Eq.~\ref{eq:sinc_window_filter}), where rapidly oscillating modes average to zero, while slowly varying modes accumulate coherently. This corresponds to the following condition:
\begin{align}
\Delta_l^{(q)}(\sigma;\tau) \equiv \frac{d}{dt}\Phi_l^{(q)}(\tau;t)
\simeq
0,
\end{align}
which is equivalent to a near-resonance condition for each mode $(l,q)$.
In the long-window limit (see Eq.~\ref{eq:local_delta_limit}), the sinc kernel approaches a delta function and selects the resonant frequencies:
\begin{align}
\sigma
\to
-\sigma_l^{(q)}
=
-
\left(
l\bar n_{\rm c}
+
q\dot\omega_{\rm c}
\right).
\end{align}
The evolution equations therefore reduce to discrete sums over acceleration amplitudes, evaluated at the resonant frequencies. The resulting secular evolution equations are given explicitly in Eqs.~\ref{eq:pdot_secular_selected}--\ref{eq:omegadot_secular_selected}.  For example, the semi-latus rectum evolves as:
\begin{align}
\langle
\dot p_{\rm p}
\rangle_{\mathcal I}
&\simeq
\frac{2a_{\rm c}}{\Delta t_W}
\sqrt{\frac{p_{\rm c}}{Gm}}
\nonumber \\&\sum_l
X_l^{1,0}\,
\tilde S_{\rm c}\!\left(
-\sigma_l^{(0)};\tau
\right)
e^{i\psi_l^{(0)}(\tau)}.
\label{eq:overview_pdot_resonant}
\end{align}
Here we defined an important quantity, i.e. the residual binary phase $\psi_l^{(q)}$:
\begin{align}
    \psi_l^{(q)} = -\sigma_l^{(q)}\tau
+
lM_{\rm c}(\tau)
+
q\omega_{\rm c}(\tau), 
\end{align}
which controls for how long the linearization of the orbital phases is valid. In other words, the residual binary phase is always approximately stationary within an adiabatic window. However, it starts to rapidly accumulate as the binary evolves via radiation reaction, describing how the orbital frequencies are changing in time.

\subsubsection{Mean anomaly response}
Unlike the other orbital elements, the mean anomaly contains a non-vanishing Keplerian contribution even in the absence of environmental forcing. Recall its LPE:
\begin{align}
\dot M
=
\bar n
-
\frac{2r}{\bar n a^2}R
+
\sqrt{1-e^2}
\left(
\dot\omega
+
\cos\iota\,\dot\Omega
\right),
\label{eq:overview_mean_anomaly_lpe}
\end{align}
where $\bar n=\sqrt{Gm/a^3}$ is the instantaneous osculating mean motion and $m$ is the total mass. The first term describes the intrinsic Keplerian orbital sweep, the second term is a direct radial forcing contribution, while the remaining terms arise from re-orientations of the osculating ellipse. As shown in Appendix \ref{Appendix:Resonant_response:Mean_anomaly}, linearizing around the carrier orbit yields the following perturbation equation:
\begin{align}
\dot M_{\rm p}
=
\dot M^{\rm Kep}
+
\dot M^{\rm dir}
+
\dot M^{\rm geom},
\label{eq:overview_mean_response}
\end{align}
where we defined: 
\begin{align}
\dot M^{\rm Kep}
&\equiv
-
\frac{3}{2}
\bar n_{\rm c}
\frac{a_{\rm p}}{a_{\rm c}},\\
\dot M^{\rm dir}
&\equiv
-
\frac{2r}{\bar n_{\rm c}a_{\rm c}^2}
R_{\rm c},\\
\dot M^{\rm geom}
&\equiv
\sqrt{1-e_{\rm c}^2}
\left(
\dot\omega_{\rm p}
+
\cos\iota_{\rm c}\dot\Omega_{\rm p}
\right).
\label{eq:mean_anomaly_force_split}
\end{align}
and each term corresponds directly to the LPE above.
Averaging the mean anomaly evolution over each adiabatic window, the resulting secular response naturally separates into three physically distinct contributions:
\begin{align}
\langle
\dot M_{\rm p}
\rangle_{\mathcal I}
=
\langle
\dot M_{\rm p}
\rangle_{\mathcal I}^{\rm kep}
+
\langle
\dot M_{\rm p}
\rangle_{\mathcal I}^{\rm dir}
+
\langle
\dot M_{\rm p}
\rangle_{\mathcal I}^{\rm geom}
.
\label{eq:overview_phase_split}
\end{align}
The first contribution describes the accumulated shift of the orbital phase due to the long-term evolution of the semi-major axis and results from two integrals over the full perturbation history of the binary. The second contribution corresponds to perturbations to the orbital velocity at fixed orbital geometry. The simplest realization of this is given by a constant radial force altering the centrifugal balance of the orbit, and therefore modifying the angular volocity at constant radius. The third contribution corresponds to geometric re-orientations of the perturbed orbit relative to the carrier orbit, i.e. the orbit geometry being twisted around by the perturbation.
\subsubsection{Integration over resonances}

The final step is to connect the local, window-averaged response to the long-term evolution of the binary. We require an efficient method to evaluate the evolution equations for cases where the environmental phase is changing rapidly and the resonances are sharp. As shown in Appendix \ref{Appendix:Resonant_response:radiation_reaction}, this can be achieved using the stationary-phase approximation (SPA). For a given mode $(l,q)$, the relevant phase is the combination of the binary residual phase $\psi_l^{(q)}(\tau)$ together with the phase of the environmental forcing spectrum. The coherency of the two determines stationarity, given by the condition:
\begin{align}\label{eq:gamma_l-spa}
\Gamma_l^{(q)}(\tau)
\equiv
\frac{d}{d\tau}
\psi_{\rm env}(\tau)
+
\frac{d}{d\tau}
\psi_l^{(q)}(\tau) \simeq 0,
\end{align}
with associated coherence time for each stationary point:
\begin{align}\label{eq:coherence-time-spa}
t_{l,\, {\rm sp}}^{\rm coh}
=
\left(\left|
\frac{d\Gamma_l^{(q)}}{d\tau}
\right|\bigg\rvert_{\tau=\tau_{\rm sp}}\right)^{-1/2}.
\end{align}

Assuming isolated and non-degenerate stationary points, the accumulated kick from each resonant crossing is obtained from the SPA integral, where e.g.:
\begin{align}
\Delta p_{\rm sp}
&=
\int
\langle
\dot p_{\rm p}
\rangle_{\mathcal I}
\,d\tau
\\&\propto
\sum_{l,q}\sum_{\rm sp}
\sqrt{2\pi}\,
X_l^{0,1} \tilde S_{\rm c}(-\sigma_l;\tau)\,
t_{l,\,{\rm sp}}^{\rm coh}\nonumber \\&\times
\exp\!\left[i \psi_{\rm tot}^{\rm SPA}
\right] \bigg\rvert_{\tau=\tau_{\rm sp}},
\end{align}
where for each $(l,q)$ mode:
\begin{align}
 \psi_{\rm tot}^{\rm SPA}=   \psi_{\rm env}
+
\psi_l^{(q)}
+
\frac{\pi}{4}
\mathrm{sgn}
\!\left(
\frac{d\Gamma_l^{(q)}}{d\tau}
\right).
\end{align}
Thus, the overall evolution is approximated as a sum over all resonant kicks at the corresponding stationary points, for each relevant mode. The amplitude of the kicks depend on the strength of the magnitude of the acceleration amplitudes, the relative phase between the binary and the environmental forcing, and the coupling given by Hansen coefficients. In practice, the SPA jumps can be implemented as linear or Gaussian ramps that accumulate over one coherence timescale rather than discontinuities, which better approximates the result of direct integration.

\subsection{Final ODEs}

\begin{table}
\centering
\begin{tabular}{ll}
\hline
\textbf{Term} & \textbf{Reference} \\
\hline

$\langle \dot p_{\rm p} \rangle_{\mathcal I}^{\rm res}$
& Eq.~\ref{eq:pdot_secular_selected}
\\

$\langle \dot e_{\rm p} \rangle_{\mathcal I}^{\rm res}$
& Eq.~\ref{eq:edot_secular_selected}
\\

$\langle \dot\iota_{\rm p} \rangle_{\mathcal I}^{\rm res}$
& Eq.~\ref{eq:iotadot_secular_selected}
\\

$\langle \dot\Omega_{\rm p} \rangle_{\mathcal I}^{\rm res}$
& Eq.~\ref{eq:Omegadot_secular_selected}
\\

$\langle \dot\omega_{\rm p} \rangle_{\mathcal I}^{\rm res}$
& Eq.~\ref{eq:omegadot_secular_selected}
\\

$\langle \dot M_{\rm p} \rangle_{\mathcal I}^{\rm dir}$
& Eq.~\ref{eq:mean_direct_avg}
\\

$\langle \dot M_{\rm p} \rangle_{\mathcal I}^{\rm geom}$
& Eq.~\ref{eq:mean_geometric_avg}
\\

$\bar n(p,e)$
& Eq.~\ref{eq:mean_anomaly_lpe}
\\

$\dot p_{\rm rr}(p,e)$
& Vacuum RR model
\\

$\dot e_{\rm rr}(p,e)$
& Vacuum RR model
\\

$\dot\iota_{\rm vac}(p,e)$
& Vacuum precession model
\\

$\dot\Omega_{\rm vac}(p,e)$
& Vacuum precession model
\\

$\dot\omega_{\rm vac}(p,e)$
& Vacuum precession model
\\
\hline
\end{tabular}
\caption{Look-up table for the final ODEs. The vacuum contributions are supplied by the user.}
\label{tab:ode_lookup}
\end{table}
In the previous sections we developed an efficient framework for computing the perturbations to the orbital elements induced by external forces. Over the course of a full inspiral however, these perturbations feed back into the radiation-reaction and relativistic precession terms, producing additional Jacobian corrections to the orbital evolution. Additionally, the perturbations to the orbital phases can reach $\mathcal{O}(1)$, which breaks the linear perturbation theory use to compute the response. A simple and self-consistent way to account for these effects (see Appendix \ref{Appendix:Resonant_response:radiation_reaction} for more detail) is to augment the LPE with the orbit-averaged radiation-reaction and precession terms, add the perturbative corrections derived above, and then integrate the resulting system of equations directly for the full orbital parameters $x = x_{\rm c} + x_{\rm p}$. The final outcome of our derivation is therefore a coupled set of evolution equations for the full orbital elements:

\begin{align}
\dot p
&=
\dot p_{\rm rr}(p,e)
+
\langle
\dot p_{\rm p}
\rangle_{\mathcal I}
,
\\
\dot e
&=
\dot e_{\rm rr}(p,e)
+
\langle
\dot e_{\rm p}
\rangle_{\mathcal I},
\\
\dot\iota
&=
\dot\iota_{\rm vac}(p,e)
+
\langle
\dot\iota_{\rm p}
\rangle_{\mathcal I},
\\
\dot\Omega
&=
\dot\Omega_{\rm vac}(p,e)
+
\langle
\dot\Omega_{\rm p}
\rangle_{\mathcal I},
\\
\dot\omega
&=
\dot\omega_{\rm vac}(p,e)
+
\langle
\dot\omega_{\rm p}
\rangle_{\mathcal I},
\\
\dot M
&=
\bar n(p,e)
+
\langle
\dot M_{\rm p}
\rangle_{\mathcal I}^{\rm dir}
+
\langle
\dot M_{\rm p}
\rangle_{\mathcal I}^{\rm geom}.
\end{align}
The explicit expressions for each contribution are collected
in Table~\ref{tab:ode_lookup}. The vacuum radiation reaction and precession terms can appropriately supplied given the application. The individual perturbations of the orbital parameters can be found simply by subtracting the carrier orbit, i.e. the orbit integrated without external perturbations, to the result of the full integration. The ODEs can also be supplied with spin evolution equations if relevant. Finally, note that for direct applications to GW observables, $\dot{p}_{rr}$ and $\dot{e}_{rr}$ must be re-expanded in terms of the waveform
phase to account for the frequency shifts arising from the conservative effect of the
environment on the orbit.

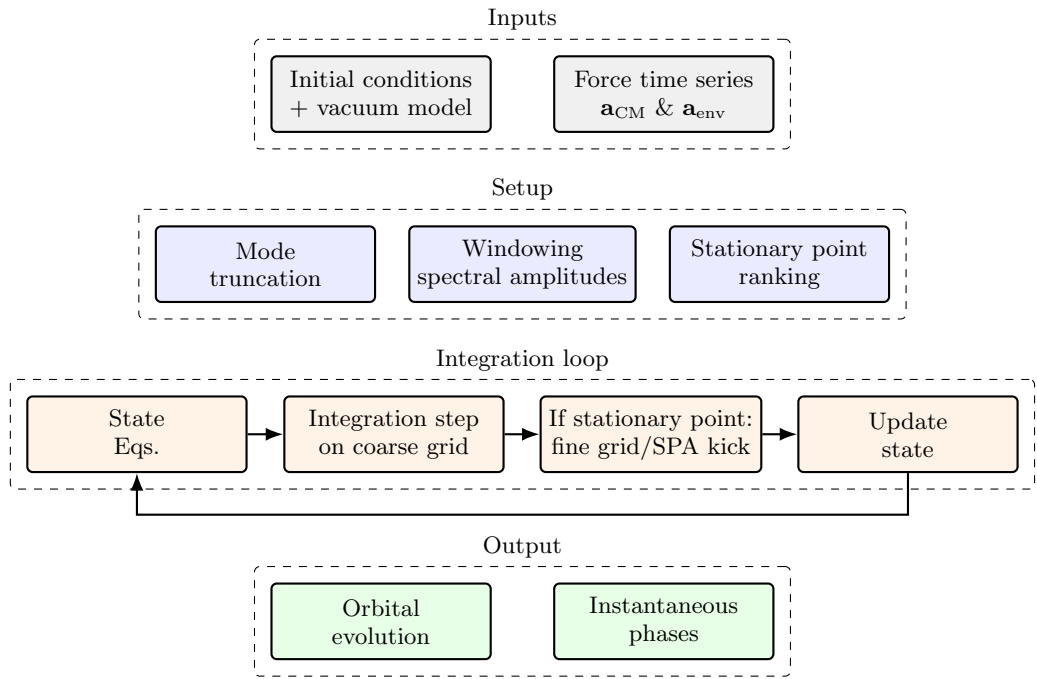
\begin{figure*}[t]
\centering
\begin{tikzpicture}[
    font=\small,
    >=Latex,
    box/.style={
        draw,
        rounded corners=2pt,
        thick,
        minimum width=2.9cm,
        minimum height=1.0cm,
        align=center
    },
    input/.style={box, fill=gray!12},
    prep/.style={box, fill=blue!8},
    loop/.style={box, fill=orange!10},
    output/.style={box, fill=green!10},
    arrow/.style={->, thick}
]

\def\dx{3.4}
\def\dy{2.25}


\node[input] (ics)   at (-0.55*\dx,0) {Initial conditions\\+ vacuum model};
\node[input] (force) at (0.55*\dx,0)  {Force time series\\$\mathbf a_{\rm CM}$ \& $\mathbf a_{\rm env}$};

\node[
    draw,
    dashed,
    rounded corners=2pt,
    inner sep=6pt,
    fit=(ics)(force),
    label={[font=\small]above:Inputs}
] (inputbox) {};


\node[prep] (window)   at (-1*\dx,-\dy) {Mode\\truncation};
\node[prep] (spectrum) at (0,-\dy)      {Windowing\\spectral amplitudes};
\node[prep] (truncate) at (1*\dx,-\dy)  {Stationary point\\
ranking};

\node[
    draw,
    dashed,
    rounded corners=2pt,
    inner sep=6pt,
    fit=(window)(spectrum)(truncate),
    label={[font=\small]above:Setup}
] (setupbox) {};


\node[loop] (state)  at (-1.5*\dx,-2*\dy) {State\\Eqs.};
\node[loop] (step)   at (-0.5*\dx,-2*\dy) {Integration step\\on coarse grid};
\node[loop] (spa)    at (0.5*\dx,-2*\dy)  {If stationary point:\\ fine grid/SPA kick};
\node[loop] (update) at (1.5*\dx,-2*\dy)  {Update\\state};

\draw[arrow] (state) -- (step);
\draw[arrow] (step) -- (spa);
\draw[arrow] (spa) -- (update);

\draw[arrow]
    (update.south) -- ++(0,-0.55) -| (state.south);

\node[
    draw,
    dashed,
    rounded corners=2pt,
    inner sep=6pt,
    fit=(state)(step)(spa)(update),
    label={[font=\small]above:Integration loop}
] (loopbox) {};


\node[output] (evolution) at (-0.55*\dx,-3.1*\dy) {Orbital\\evolution};
\node[output] (dephase)   at (0.55*\dx,-3.1*\dy)  {Instantaneous\\phases};

\node[
    draw,
    dashed,
    rounded corners=2pt,
    inner sep=6pt,
    fit=(evolution)(dephase),
    label={[font=\small]above:Output}
] (outputbox) {};

\end{tikzpicture}

\caption{Practical workflow of the LFH framework. The inputs specify the binary initial conditions and the time series of the perturbing accelerations evaluated on the vacuum carrier orbit (including the center of mass acceleration). The setup stage pre-processes the force into a truncated set of dominant spectral modes and identifies the most relevant stationary points. The integration loop evolves the smooth ODE system and applies stationary-phase kicks or fine integration when resonant crossings are encountered. The outputs are the full orbital evolution and phases phases, which can be supplied to a waveform generator.}
\label{fig:LFH_practical_workflow}
\end{figure*}
\subsection{Line of sight acceleration}
\label{sec:instantaneous_gw_phase}

While the orbital dynamics are driven by the difference of forces on the individual binary components:
\begin{align}
\mathbf a_{\rm env}
=
\mathbf a_1-\mathbf a_2,
\end{align}
the acceleration of the binary center of mass is:
\begin{align}
\mathbf a_{\rm CM}
=
\frac{
m_1\mathbf a_1
+
m_2\mathbf a_2
}{
m_1+m_2
}.
\label{eq:cm_acceleration}
\end{align}
This center-of-mass acceleration produces an observer-frame timing correction. In-keeping with the rest of the LFH framework, we average the center of mass acceleration over the local window:
\begin{align}
\left\langle
\mathbf a_{\rm CM}
\right\rangle_{\mathcal I}(\tau)
=
\frac{1}{\Delta t_W}
\int_{\mathcal I_\tau}
\mathbf a_{\rm CM}(t')\,dt' .
\label{eq:cm_acceleration_average}
\end{align}

Defining $\hat{\mathbf N}$ as the unit vector from the source to the observer, the corresponding Rømer delay is:
\begin{align}
\Delta t_{\rm R}(\tau)
=
-\frac{1}{c}
\hat{\mathbf N}\cdot
\int_{\tau_0}^{\tau}
d\tau'
\int_{\tau_0}^{\tau'}
\left\langle
\mathbf a_{\rm CM}
\right\rangle_{\mathcal I}(\tau'')
\,d\tau'' ,
\label{eq:roemer_delay_acceleration}
\end{align}
up to constant position and velocity terms, which are degenerate with an overall time and frequency shift. To first order in small Rømer delays, the observed phase $\phi^{\rm obs}$ of the GW will be delayed:
\begin{align}
\phi^{\rm obs}(\tau)
=
\phi(\tau)
-
\dot \phi(\tau)\,
\Delta t_{\rm R}(\tau),
\label{eq:observed_phase_roemer}
\end{align}
where $\phi\equiv\nu+\omega$ is the planar phase of the binary.

\subsection{Frequency domain phases for GWs}
The fastest and most efficient GW waveforms are represented in the frequency-domain \cite{1994cutler,purrer}. In order to enable easy implementation in such waveforms, time-domain GW phase shifts need to be converted into frequency-domain GW phase shifts. In the time domain, eccentric and precessing waveforms are given as a sum over harmonics:
\begin{align}
    h(t)=\sum_{ij}h_{ij}(t)\,e^{i\phi_{ij}},
\end{align}
where we neglected projection effects. The phase $\phi_{a,b}$ of individual GW harmonics can be written in terms of orbital angles:
\begin{equation}
    \phi_{ij}(\tau) = a M(\tau) + b \omega(\tau).
\end{equation}
Thus the time-domain phase shift of each harmonic simply becomes:
\begin{equation}
    \delta\phi_{ij} = i  M_{\rm p} + j \omega_{\rm p} .
\end{equation}
Transforming to the frequency domain involves applying the SPA for each individual GW harmonic, with the condition that the frequency of each harmonic is slowly evolving, i.e. $\ddot{\phi}_{\ell,m} \ll \dot{\phi}_{\ell,m}^2$.
Then one can rely on the simple relation between the time- and frequency-domain phase shifts, valid up to first-order in the perturbations (see e.g. \cite{pedo}) to obtain:
\begin{equation}
    \delta\Psi_{ij}(f_{ij}) = -\delta\phi_{ij}(\tau(f_{ij})),
\end{equation}
where $\Psi_{i,i}(f)$ is the frequency domain GW dephasing, and:
\begin{equation}
    f_{ij}(t) = i \bar n + j \dot{\omega},
\end{equation}
and finally  $\tau(f_{ij})$ is the stationary time corresponding to when $\dot \phi_{ij}(\tau) = 2\pi f_{ij}$ in the carrier phase.

\begin{figure*}[t]
    \centering
    \includegraphics[width=1\linewidth]{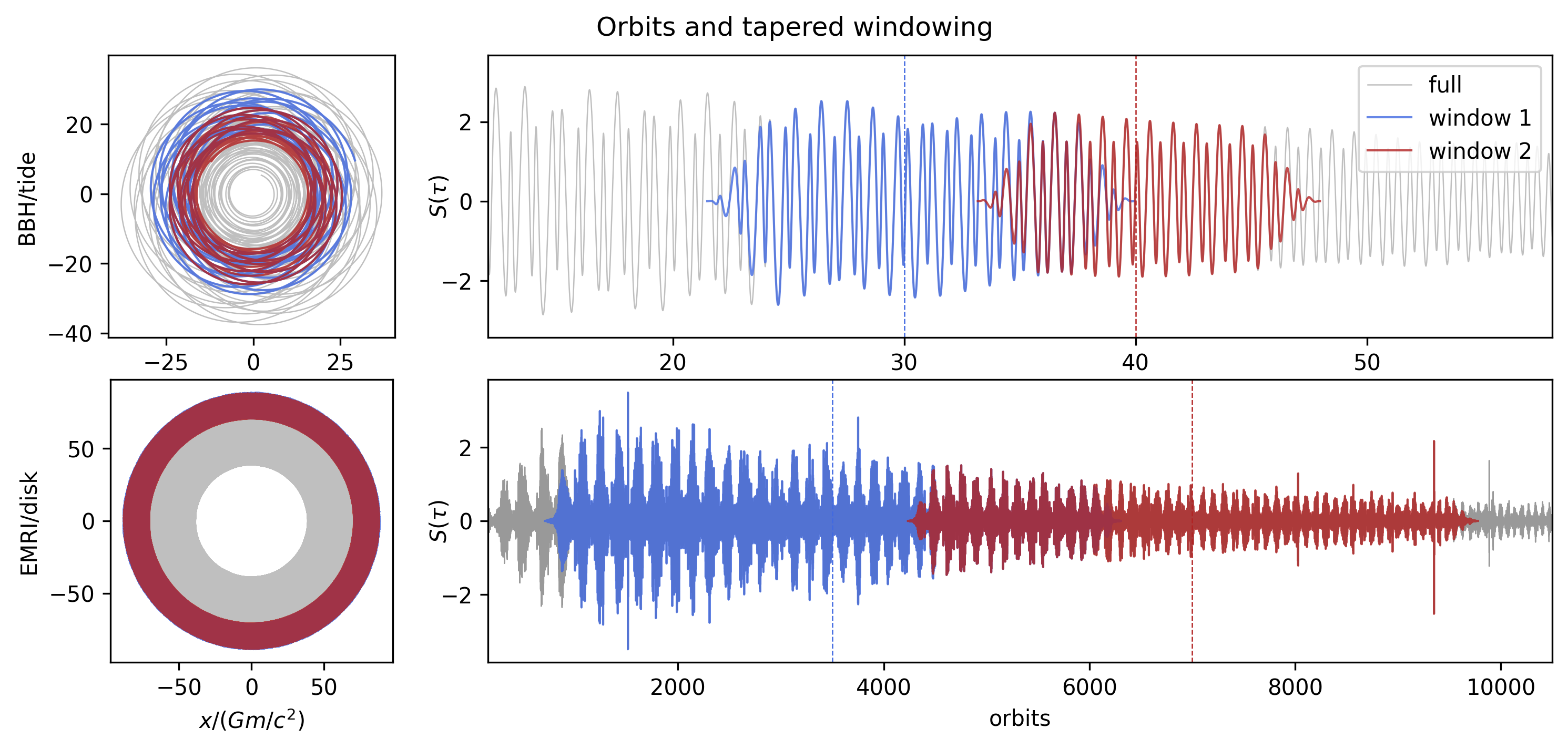}
    \caption{Illustration of the windowing step for two example orbits and perturbation time series. Top panel: Orbit and tangential acceleration for a $10\,{\rm M}_\odot+10\,{\rm M}_\odot$ binary embedded in a rotating tidal field starting from an orbital frequency of 10 Hz and $e = 0.3$. Bottom panel: Orbit and tangential acceleration of a $100\,{\rm M}_\odot$ EMRI orbiting a $10^5\,{\rm M}_\odot$ BH, starting from an orbital frequency of $4\times 10^{-4}$ Hz and $e = 0.03$, interacting with an accretion disk (extracted from numerical simulations detailed in \cite{Derdzinksi:2021}).  The overlayed colored segments indicate subsequent and overlapping tapered windows, as required for the LFH framework. The dashed vertical lines denote the chosen window centers, two for this illustration.}
    \label{fig:orbit_torque}
\end{figure*}
\begin{figure}
    \centering
    \includegraphics[width=1\linewidth]{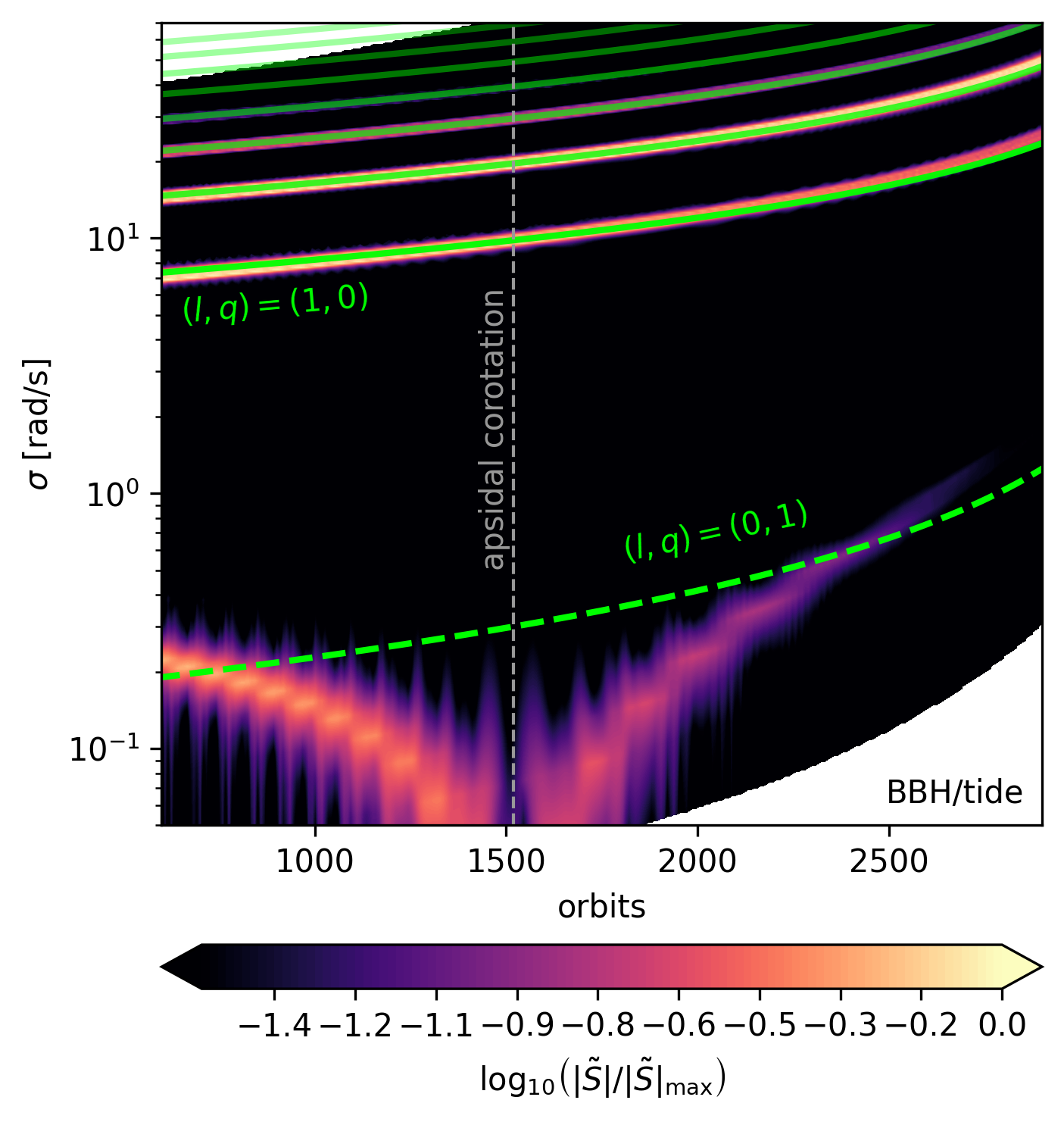}
    \caption{Spectrogram of the tangential acceleration for a $10\,{\rm M}_\odot+10\,{\rm M}_\odot$ binary embedded in a rotating tidal field starting from an orbital frequency of 1 Hz and $e = 0.3$. The contours show the power in the windowed Fourier transform of $S_{\rm c}(\tau)$, centered at $\tau$ (rescaled to orbits). The green lines denote the evolution of the binary frequencies $l \bar n_{\rm c} +q\dot \omega_{\rm c} $ as the inspiral progresses. Modes that intersect regions of high power have the potential to create large resonant kicks. Similarly, times at which the perturbations becomes ``DC'', or zero-frequency, always create an orbital response. Here this happens at around orbit 1520, highlighted by the vertical gray line. The white regions are undefined due to the finite duration of the averaging windows. }
    \label{fig:spectrogram}
\end{figure}
\section{The LFH framework in practice}
\label{sec:practice}
In this section, we outline the practical workflow of the LFH framework. To illustrate the steps concretely, we use two perturbative force time series motivated by important astrophysical applications. The first corresponds to a stellar-mass binary embedded in a rotating tidal field, while the second uses forces extracted from a numerical simulation of an EMRI interacting with an accretion disc. Here, these systems are used only as illustrative examples, showcasing how the same formalism can treat highly structured, analytical perturbations as well as broadband forces extracted from numerical simulations. More context for astrophysical applications is provided in section \ref{sec:discussion}. An overview of the practical workflow is provided in Fig. \ref{fig:LFH_practical_workflow}.

\begin{figure*}
    \centering
    \includegraphics[width=1\linewidth]{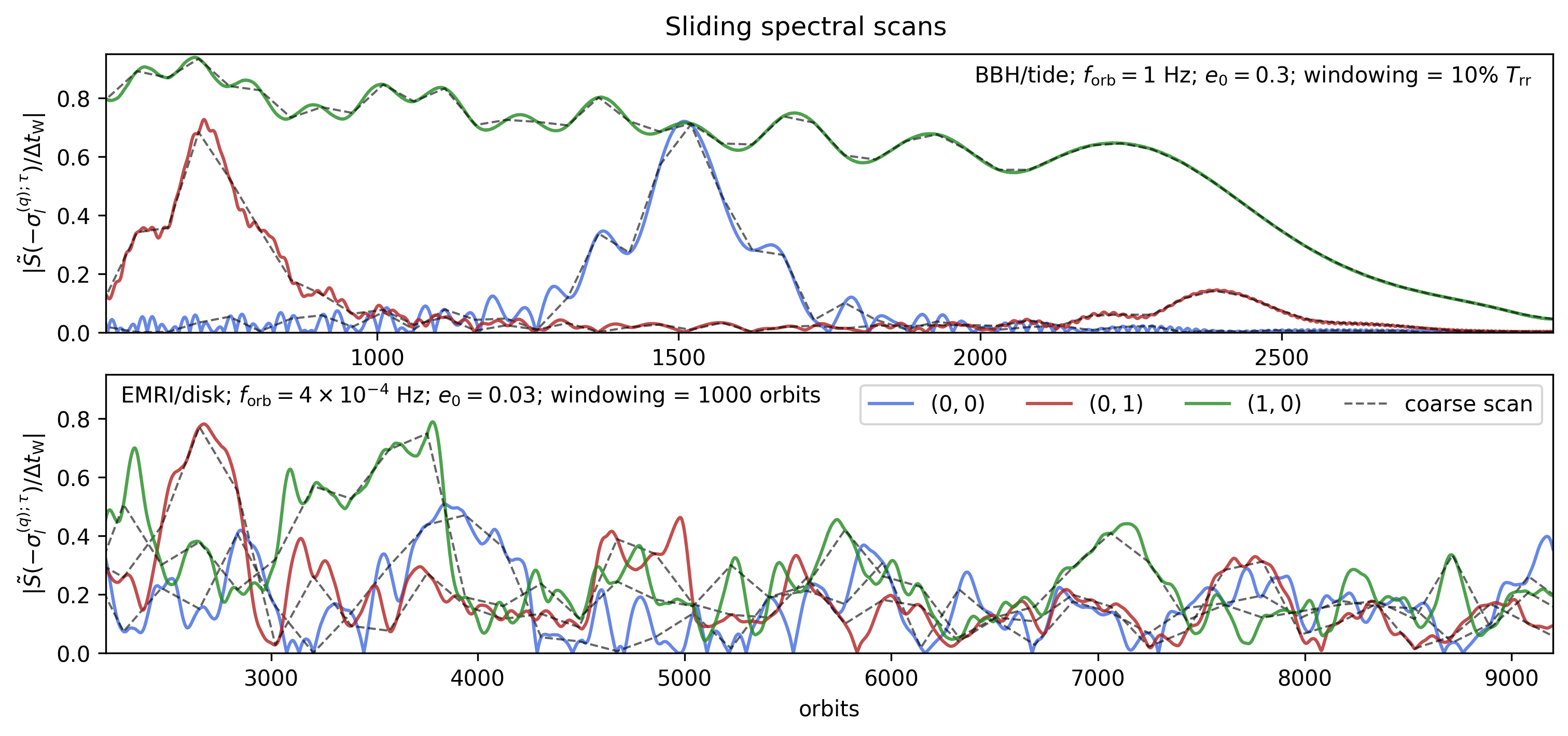}
    \caption{Scan of tangential acceleration Fourier amplitudes evaluated at the modes $(l,q)=(0,0)$, $(0,1)$ and $(1,0)$. The smooth colored lines are computed with a finely separated grid of 5000 averaging windows, while the grey dashed lines indicate the same analysis performed on only 50 window centers. Note how the major features are preserved even in a coarse computation. Top panel: Results for a $10\,{\rm M}_\odot+10\,{\rm M}_\odot$ binary embedded in a rotating tidal field starting from an initial orbital frequency of 1 Hz and $e = 0.3$. Note the distinct apsidal co-rotation resonances in the   $(0,0)$ and $(0,1)$ modes, as well as the large amount of power for the $(l,0)$ mode oscillating at the orbital frequency. Bottom panel: Results for a $100\,{\rm M}_\odot$ EMRI orbiting a $10^5\,{\rm M}_\odot$ BH, starting at $f_{\rm GW}=2\times10^{-4}$ Hz and $e = 0.03$, interacting with an accretion disk (extracted from numerical simulations detailed in \cite{Derdzinksi:2021}). Note how different modes dominate at different stages of the inspiral.}
    \label{fig:sliding_scans}
\end{figure*}
\begin{figure*}
    \centering
    \includegraphics[width=1\linewidth]{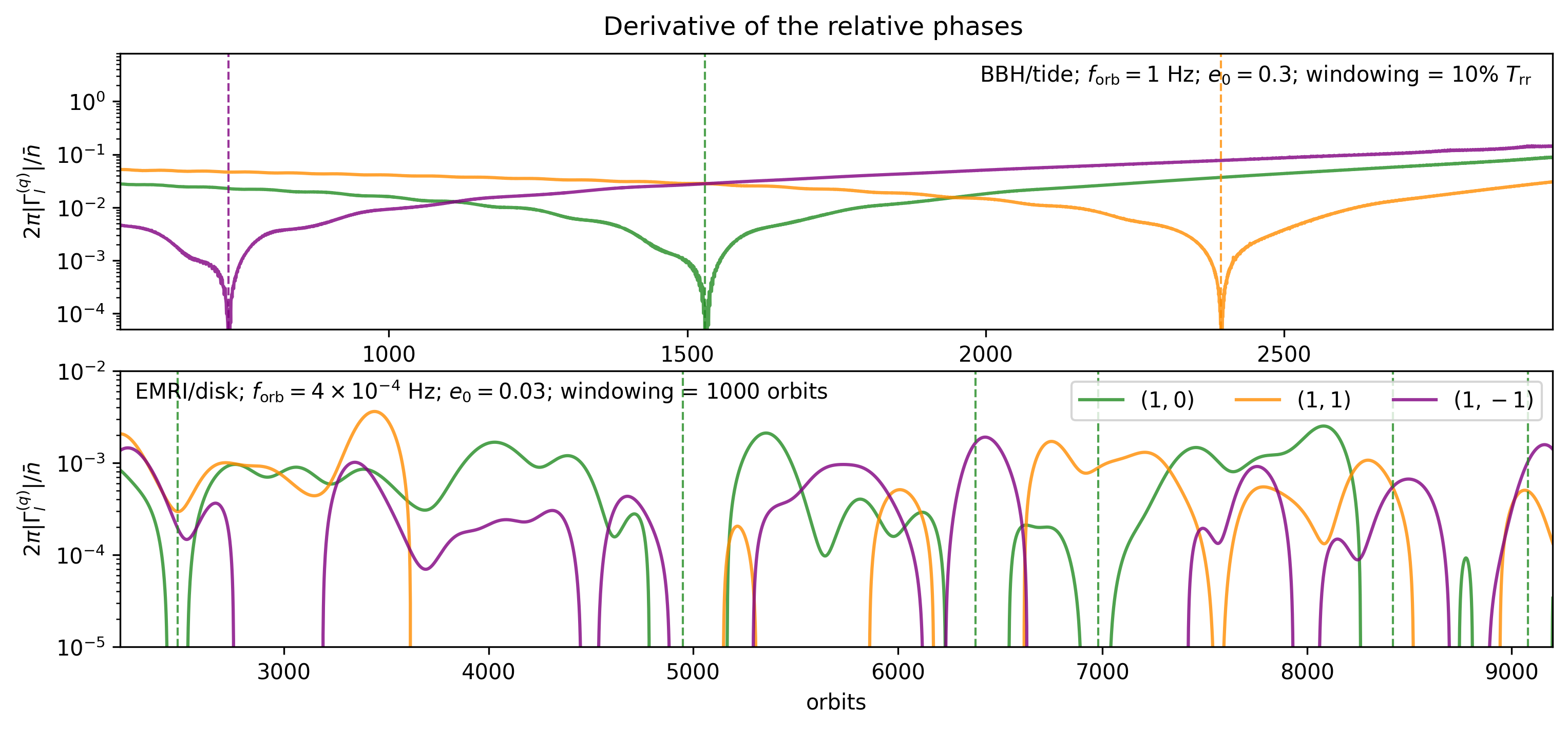}
    \caption{
Derivative of the relative phase of the tangential acceleration for the modes $(1,0)$, $(1,1)$, and $(1,-1)$, computed using a dense grid of 5000 overlapping adiabatic windows. Top panel: normalized relative phase derivative, $2\pi |\Gamma_l^{(q)}|/\bar n$, for a $10\,{\rm M}_\odot+10\,{\rm M}_\odot$ binary embedded in a rotating tidal field, with initial orbital frequency $f_{\rm orb}=1\,{\rm Hz}$ and eccentricity $e_0=0.3$. Smaller values correspond to longer phase coherence. Vertical dashed lines indicate candidate stationary points where $\Gamma_l^{(q)}\simeq0$, leading to enhanced resonant coupling. The $(0,0)$ mode is omitted. Bottom panel: same as above for a $100\,{\rm M}_\odot$ compact object orbiting a $10^5\,{\rm M}_\odot$ black hole with initial eccentricity $e_0=0.03$, interacting with an accretion disc. The force time series is extracted from the numerical simulations of \cite{Derdzinksi:2021}. Here we see how several modes remain approximately phase coherent over intervals of $10^2$--$10^3$ orbital cycles, resulting in a large number of stationary points (highlighted as vertical lines only for the $(1,0)$ mode). Note that here the EMRI results have some additional smoothing to add clarity.}
    \label{fig:phases}
\end{figure*}
\subsection{Inputs, perturbations and mode truncation}
The starting point of the LFH framework is a binary specified by initial orbital elements:
\begin{align}
(p_0,e_0,\omega_0,\Omega_0,\iota_0,M_0),
\end{align}
together with a prescription for the vacuum inspiral evolution and precession rates. For the purposes of this work, we adopt the leading-order post-Newtonian expressions for both radiation reaction and conservative precession \cite{1938einstein,peters1964}. In addition, one requires a perturbing acceleration time series evaluated on carrier binary trajectory. For our first example, we consider an analytical perturbing acceleration corresponding to a tidal field whose principal axis rotates with frequency $\Omega_{\rm env}$. For a planar binary configuration, the acceleration components are:
\begin{align}
\label{eq:tidal_fieldforces}
R(\tau)
&=
\mathcal T\, r(\nu)
\left[
\frac{1}{2}
+
\frac{3}{2}
\cos\left(
2(\omega+\nu-\tau\,\Omega_{\rm env})
\right)
\right],
\\
S(\tau)
&=
-\frac{3}{2}
\mathcal T\, r(\nu)
\sin\left(
2(\omega+\nu-\tau\,\Omega_{\rm env})
\right),
\end{align}
where $\mathcal T$ sets the strength of the tidal perturbation. This represents a highly structured perturbation that oscillates at a few localized frequencies. Throughout, we will use representative binary parameters for a LVK or CE/ET source, consisting of a $10$ M$_{\odot}$ + 10 M$_{\odot}$ binary with an initial orbital frequency of either 10 Hz or 1 Hz, respectively. We will choose values for $\Omega_{\rm ext} \sim \dot \omega$ in the early inspiral, to showcase the emergence of an apsidal co-rotation resonance. For our second example, we consider the torque time series experienced by an EMRI embedded in an accretion disk, extracted from the hydrodynamic simulations of \cite{Derdzinksi:2021} \footnote{The hydro-dynamical simulations referred to here were performed for a circular EMRI.  We will assume that the numerical results for the torque time series are representative for small eccentricities of $\lesssim 0.03$, given the same simulation setup.}. We choose representative mass scales of 100 M$_{\odot}$ and $10^5$ M$_{\odot}$, with an initial orbital frequency of $4\times 10^{-4}$ Hz, such that the 12000 orbits resolved in the simulation occur in approximately one year of observation. Example orbital trajectories and force time series for both systems are shown in Fig.~\ref{fig:orbit_torque}. The center of mass force $\mathbf{a_{\rm CM}}$ can also be treated as noted in section \ref{sec:framework_overview}, but is not the focus here.

The next step is to determine how many orbital harmonics must be retained in the Fourier--Hansen expansion. In practice, this truncation is controlled primarily by the eccentricity dependence of the Hansen coefficients. At leading order, they scale approximately as (see Appendix \ref{Appendix:Hansen_coeff}):
\begin{align}
X_l^{j,k}(e)
\sim
e^{|l-k|},
\label{eq:hansen_scaling_practical}
\end{align}
where in the relevant LPE one has $|k|\leq1$. Consequently, near-circular binaries are dominated by the lowest harmonics, while highly eccentric systems require a broader selection of modes. Throughout this work we adopt a representative initial eccentricity of $e=0.3$ for the binary BH example and $e=0.03$ for the EMRI and retain harmonics up to $l=1$ or $l=2$ depending on the application. In practice the truncation depends both on the desired numerical precision and on the detailed structure of the perturbations spectrum. A lot of information can be gained at a glance by computing a spectrogram of the perturbing accelerations, as is shown in Fig.~\ref{fig:spectrogram} for the case of the binary in a tidal field. Note how the first few orbital harmonics $(l,0)$ track contours of with large ``power'', highlighting that resonances in those modes would produce large kicks in the orbital elements. Note additionally the presence of a co-rotation resonance in the $(l,q)=(0,0)$ mode. This is realised when power that is initially at $\sim 10^{-1}$ Hz suddenly becomes ``DC'', i.e. goes to zero-frequency. Then, for this period of time, the $(0,0)$ mode of the binary responds to this environmental forcing, producing a kick in the orbital evolution. Many corresponding features of this spectrogram can also be seen in the top panel of Fig. \ref{fig:sliding_scans}.

\subsection{Windowing and spectral decomposition}
The next step is to choose an initial grid in $\tau$:
\begin{align}
    {\rm Integration \, grid} = [\tau_0,\, \tau_1,...,\,\tau_n],
\end{align}
which determines the time resolution of the integration. Then, the time series of the perturbing acceleration is separated into overlapping adiabatic windows with lengths $\Delta t_{W;\, i}$ satisfying:
\begin{align}
T_{\rm orb}(\tau_i)
\ll
\Delta t_{W;\, i}
\ll
T_{\rm rr}(\tau_i).
\end{align}
Each window is smoothly tapered (e.g. with a Tukey window \cite{1958BSTJ...37..485B}, see Fig. \ref{fig:orbit_torque}) to suppress edge effects. Then, for each window, the projected force components are Fourier transformed and the resulting local spectra are evaluated at the resonant frequencies selected by the Fourier--Hansen expansion (see section \ref{sec:framework_overview}):
\begin{align}
\sigma
\rightarrow
-\sigma_l^{(q)}
=
-
\left(
l\bar n
+
q\dot\omega
\right)
\end{align}
This step extracts the force amplitudes that can couple coherently to the orbital harmonics within each adiabatic window. Fig.~\ref{fig:sliding_scans} shows the resulting spectral amplitudes for our two worked examples, evaluated at selected resonant frequencies as the window is moved along the inspiral. The figure illustrates how the environmental coupling evolves in time, and how different resonant channels can emerge, fade, or exchange dominance as the binary evolves. The chosen width of the adiabatic windows controls the trade-off between time and frequency resolution. Shorter windows provide better localization of transient features, while longer windows yield smoother spectra and improved frequency resolution. The figure also compares coarse and fine grids of window centers $\tau_i$. Fine grids resolve rapid variations in the local spectral content, whereas coarser grids can substantially reduce the computational cost of the subsequent orbital integration, while also retaining the dominant features of the forcing. In practice, the optimal choice of window length and sampling density depends on the characteristic timescales of the perturbation and on the accuracy requirements of the waveform model. We do not investigate these choices systematically here; a detailed discussion, together with performance benchmarks and implementation details, will be presented in the forthcoming \texttt{AstroWaveforms} paper.

\subsection{Identification of the stationary points}
The next step is to identify the stationary points at which the forcing and binary configuration evolve coherently, and rank them according to their expected impact on the dynamics. For each mode $(l,q)$ of each force spectrum, we construct the relative phase:
\begin{align}\label{eq:chi_force_spectrum_def}
\psi_{\rm tot}^{l,q} \equiv \psi_{\rm{env}}(\tau)+\psi_l^{(q)}(\tau),
\end{align}
where the environmental contribution is given by the argument of the relevant perturbing acceleration component $A \in (R,S,W)$:
\begin{align}
  \psi_{\rm env}(\tau) = \arg \left( \tilde A\,(-\sigma_l^{(q)};\tau) \right)\,.
\end{align}
As shown in Eq.~\eqref{eq:gamma_l-spa}, the stationary points are located at the extrema of the relative phase evolution (i.e.~$\partial_\tau \psi_{\rm{tot}}^{l,q} =0$) and their corresponding coherence times are given by the second derivative of this phase in Eq.~\eqref{eq:coherence-time-spa}.
Figure \ref{fig:phases} illustrates the coherency of the relative phase for our two worked examples, where in particular the binary in a tidal field showcases three clear stationary points in the displayed modes. Instead, for our EMRI forcing example, we observe that the relative phase has periods of coherency for $10^2$ to $10^3$ and a complex succession of stationary points with different coherence. This reflects the complex physics of gas embedded systems. In practice, the importance of a given stationary point depends on several factors, including the amplitude of the corresponding spectral mode, the Hansen weighting of the orbital harmonic, and the coherence time itself. Consequently, not all stationary points contribute equally to the orbital evolution. If required for computational efficiency, stationary points associated with weak forcing amplitudes or short coherence times can be discarded. We do not investigate such optimization strategies here, as questions of performance, truncation criteria, and computational cost will be the focus of the forthcoming \texttt{AstroWaveforms} package and its accompanying benchmarking study.

\subsection{Fully analytical procedure}
For sufficiently simple analytic perturbations, the spectral decomposition can itself be carried out analytically using an additional Hansen expansion. Here, we briefly sketch out this calculation for our worked example of a binary in a time-dependent tidal field, focusing on a few modes of the response to the tangential acceleration $S$. This serves both as an illustration on how to use the LFH framework for fully analytical systems, as well as provide some more intuition regarding the various phases and resonances mentioned in previous sections.

To start, consider that the perturbing accelerations in Eqs.~\ref{eq:tidal_fieldforces} depend only on trigonometric functions of the carrier angles $\nu_{\rm c}$ and $\omega_{\rm c}$.Within an adiabatic window, the tidal acceleration can therefore be decomposed as a sum over Hansen modes:
\begin{align}
S_{\rm c}(t)
&=
\frac{3i\mathcal T a_{\rm c}}{4}
\sum_k
X_k^{1,2}(e_{\rm c})\,
e^{i(kM_{\rm c}+2\omega_{\rm c}-2\Omega_{\rm env}t)}
\nonumber\\
&\quad-
\frac{3i\mathcal T a_{\rm c}}{4}
\sum_k
X_k^{1,-2}(e_{\rm c})\,
e^{i(kM_{\rm c}-2\omega_{\rm c}+2\Omega_{\rm env}t)}.
\end{align}
Note that here the index $k$ is distinct from the index $l$ used previously, as the latter denotes the Hansen expansion of the LPE rather than that of the perturbing acceleration. Again within each adiabatic window, the binary phases evolve linearly in time. The windowed Fourier transform can therefore be performed analytically and evaluated directly at the resonant frequencies $-\sigma_l^{(q)}$ arising from the LFH expansion. After some computation using Fourier identities, we find:
\begin{align}
\label{eq:analytical_spectral}
\tilde S(-\sigma_l^{(q)};\tau)
&=
\frac{3i\mathcal T a_{\rm c}}{4}
\sum_k
\bigg[
X_k^{1,2}(e_{\rm c})\,
e^{i\psi_k^{\rm env,2}}
\,\tilde{\mathcal W}_{\mathcal I}
\!\left(\xi_{lk,+}^{(q)}\right)
\nonumber\\
&-
X_k^{1,-2}(e_{\rm c})\,
e^{i\psi_k^{\rm env,-2}}
\,\tilde{\mathcal W}_{\mathcal I}
\!\left(\xi_{lk,-}^{(q)}\right)
\bigg],
\end{align}
where
\begin{align}
\xi_{lk,+}^{(q)}
&\equiv
-\sigma_l^{(q)}
-
\left(
k\bar n_{\rm c}
+
2\dot\omega_{\rm c}
-
2\Omega_{\rm env}
\right),
\\
\xi_{lk,-}^{(q)}
&\equiv
-\sigma_l^{(q)}
-
\left(
k\bar n_{\rm c}
-
2\dot\omega_{\rm c}
+
2\Omega_{\rm env}
\right),
\end{align}
and
\begin{align}
\psi_k^{\rm env,\pm 2}
=
kM_{\rm c}(\tau)
\pm
2\omega_{\rm c}(\tau)
\mp
2\Omega_{\rm env}\tau
+
\sigma_l^{(q)}\tau.
\end{align}
With this additional Hansen decomposition, we see that the windowed spectral amplitudes $\tilde S(-\sigma_l^{(q)};\tau)$ carry an overall phase determined both by the form of the force time series and by the state of the binary at the window centre $\tau$. This structure is directly analogous to the binary residual phase $\psi_l^{(q)}$, and in fact for this simple example the dependence on the Fourier phase cancels neatly:
\begin{align}
\label{eq:relphase_analytic}
\arg\!\left[
\tilde S(-\sigma_l^{(q)};\tau)
e^{i\psi_l^{(q)}}
\right]
&=
(k+l)M_{\rm c}
\nonumber\\
&\quad+
(q\pm2)\omega_{\rm c}
\mp
2\Omega_{\rm env}\tau.
\end{align}
This is in fact the expected behavior for a perturbation that only depends on the instantaneous binary state and not on an arbitrary choice of window center. Now, recall that $\tilde {\mathcal{W}}_{\mathcal I}$ is the frequency-domain windowing kernel corresponding to the time-domain windowing applied over the interval $\mathcal I_\tau$. For a sufficiently long averaging window $\mathcal I_\tau$, $\mathcal W_{\mathcal I}$ is sharply peaked around zero. Therefore, the windowing selects frequencies satisfying:
\begin{align}
\label{eq:analytic_conditions}
\sigma_l^{(q)}
&\simeq
k\bar n_{\rm c}
+
2\dot\omega_{\rm c}
-
2\Omega_{\rm env},
\\
\sigma_l^{(q)}
&\simeq
k\bar n_{\rm c}
-
2\dot\omega_{\rm c}
+
2\Omega_{\rm env},
\end{align}
where again, the presence of $k$ and the particular factor of 2 describe the tidal tangential acceleration. It is precisely these additional commensurability conditions between binary frequencies $\sigma_l^{(q)}$ and the frequencies of the perturbation that encode the full resonance structure of the problem. As a simple example, consider the mode with $l=0$ and $q=0$, for which the left-hand side vanishes. The commensurability condition can only be satisfied for $k=0$ and $\dot\omega_{\rm c}\simeq\Omega_{\rm env}$, corresponding precisely to apsidal co-rotation. This is visible in the spectrogram of Fig.~\ref{fig:spectrogram} and also in the amplitude enhancement seen in the sliding-window scans shown in the upper panel of Fig.~\ref{fig:sliding_scans}.

From Eq.~\ref{eq:relphase_analytic} and the conditions in Eq.~\ref{eq:analytic_conditions}, we can also identify the stationary points analytically. As an example, consider first the $(l,q)=(0,0)$ mode. In this case, $k=l=0$ and the relative phase reduces to:
\begin{align}
\psi_{\rm env}(\tau)+\psi_0^{(0)}(\tau)
=
-2\Omega_{\rm env}\tau
+
2\omega_{\rm c}(\tau).
\end{align}
We can immediately see that this phase remains coherent precisely within the adiabatic window, i.e. approximately for a radiation-reaction timescale near the stationary point defined by apsidal co-rotation:
\begin{align}
\dot\omega_{\rm c}
=
\Omega_{\rm env}.
\end{align}
This type of calculation can be repeated for any mode, in order to build up the full resonant response.

This example illustrates how an analytic force model with oscillations on the orbital timescale can be transformed into a sparse set of relevant modes. By performing the windowing analytically and studying the resulting resonance conditions together with the binary residual phases $\psi_l^{(q)}$, one can identify the dominant forcing channels, construct efficient truncated representations of the perturbation, and determine explicitly which harmonics are capable of producing stationary-phase resonances with the binary. Once the dominant modes have been selected, the resulting truncated system can be evolved directly by integrating the ODEs.

\subsection{Integration}
\begin{figure}[h!]
    \centering
    \includegraphics[width=1\linewidth]{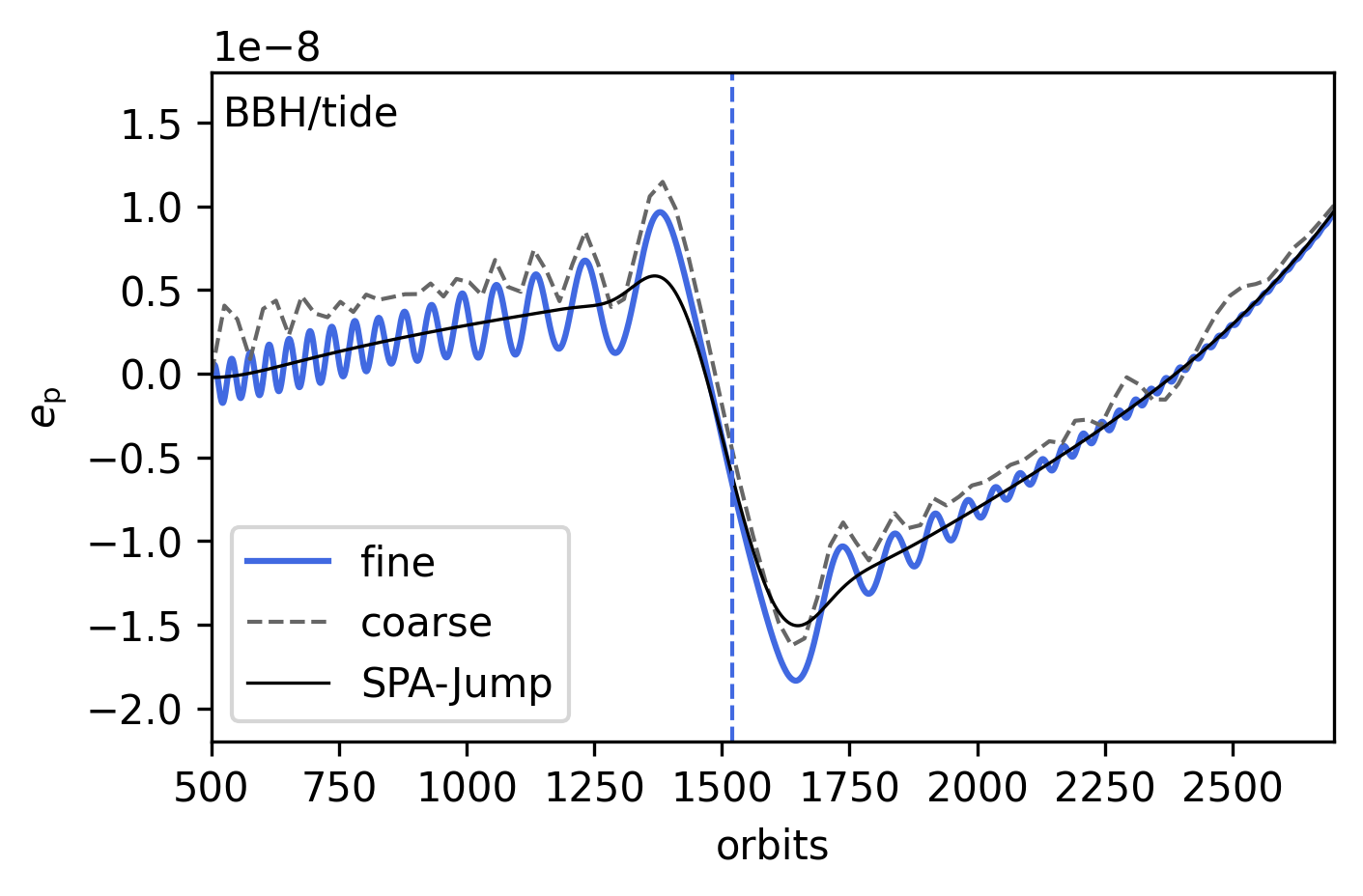}
    \caption{Illustration of the difference between integration in $\tau$ using a fine grid (5000 window centers, colored line), a coarse grid (100 window centers, dashed line) and analytical SPA jumps at resonance (solid black line). The curves represent the eccentricity perturbation for the binary in a tidal field, displayed in full in Fig.~\ref{fig:orbital_elements}. Note how the orbital evolution is recovered even with a coarse integration grid. }
    \label{fig:spa}
\end{figure}

\begin{figure}
    \centering
    \includegraphics[width=1\linewidth]{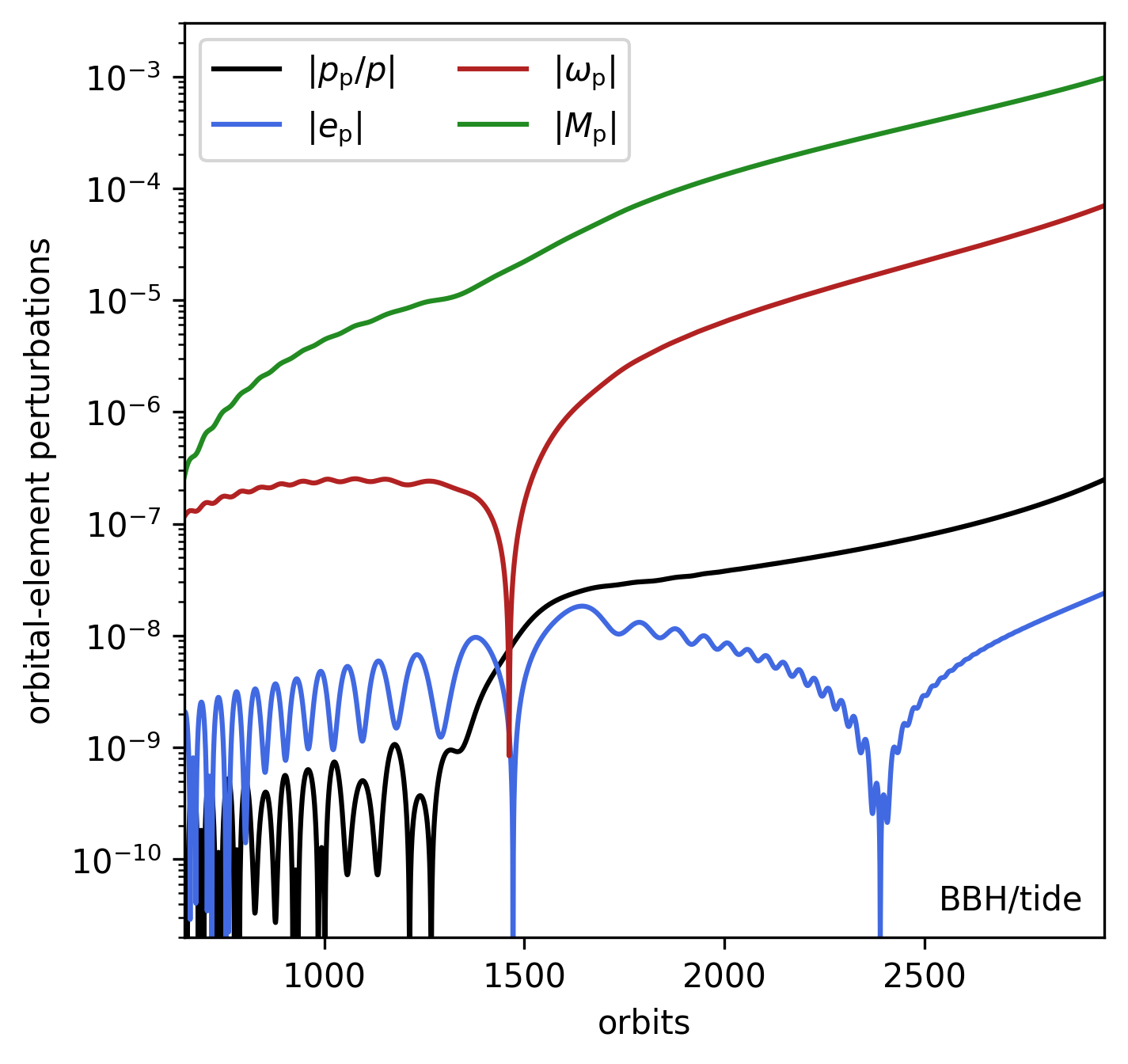}
    \caption{Integrated perturbations to the orbital elements for a $10\,M_\odot+10\,M_\odot$ binary embedded in a rotating tidal field, computed using the $l=0$, $l= \pm1$ and $q=\pm1$ modes and a fine grid of window centers.}
    \label{fig:orbital_elements}
\end{figure}

The final step is to integrate the ODEs over the desired grid of $\tau$ values. The only subtlety is performing an accurate integration over the resonances in the $(l,q)\neq(0,0)$ modes, which are mediated by the coherency of the relative phase by $\psi_{\rm env} + \psi_l^{(q)}$. Here we adopt two possible approaches: The first is simply to ensure that the resolution of the $\tau$ integration grid is sufficient to resolve the coherent forcing, i.e. that there are a few integration steps across the coherence time of the resonance in question. Note that this is still significantly more efficient than direct integration of the equations of motion, which instead require orbital timescale resolution to resolve resonances. The second approach is to use the SPA kicks discussed in Sec.~\ref{sec:framework_overview} to treat resonance crossings analytically. The difference between the two approaches is illustrated in Fig.~\ref{fig:spa}, where the analytical SPA kicks have been implemented as smooth Gaussian ramps. In practice, the SPA treatment is often significantly more efficient, since it replaces the need to resolve many orbital cycles. An automatic implementation of the SPA resonant kicks is left for the \texttt{AstroWaveforms} package and the accompanying benchmarking paper.  Figure \ref{fig:spa} also illustrates how, by using the LFH framework, accurate integration only requires a coarse integration grid. While benchmarking is left for the upcoming \texttt{Astrowaveforms} paper, we note that the resonance kick is accurately reproduced with only an integration step every $\sim 80$ orbits, as opposed to several integration steps per orbit required by direct integration of the LPE.  The full evolution of the remaining orbital elements is shown in Fig.~\ref{fig:orbital_elements}.

\section{Discussion}
\label{sec:discussion}
\subsection{Structured and stochastic environments}
Environmental effects in GW signals have most often been modeled through orbit-averaged or phenomenological prescriptions such as dynamical friction, migration torques, or constant line  accelerations. In reality, these perturbations are caused by the interaction of the binary with physical systems that posses their own dynamics and characteristic frequencies, and can therefore induce a much richer response. Moreover, astrophysical environments will often produce multiple GW environmental effects simultaneously, leading to a complex interplay of perturbations. The goal of the LFH framework is precisely to provide an efficient way to characterize the evolution and GW observables of binaries in such environments, retaining the full dynamical interaction between the relativistic binary and its surroundings. Here, we provide a brief discussion of several notable applications.

Tidal perturbations, which motivated our first worked example, arise whenever a compact binary evolves in the vicinity of a third massive body. Previous work has shown that such companions can imprint observable signatures on gravitational-wave signals through line-of-sight accelerations \citep{2017meiron,2024samsing,kai2024,kai22024,2018PhRvD..98f4012R}, secular tidal fields \citep{liu2015,2021toubiana,2023zwick,2024barandiaran,2024zwick}, and even Kozai--Lidov oscillations \citep{zeipel1910,koz62,lid62,2022chandramouli,gupta2020}. Beyond these orbit-averaged effects, the motion of the third body introduces explicitly time-dependent forcing through both the eccentric motion of the inner binary and harmonics of the outer orbital frequency, which can produce localized resonances. Our worked example considered only a single component of this interaction, corresponding to the simple case of a circular outer orbit. A much richer set of resonances emerges for eccentric outer orbits, and additional resonant channels arise once finite-size horizon effects are included \cite{pica2023,2025cocco}. Binaries with a companion are promising targets for next-generation ground-based detectors such as ET and CE, whose improved low-frequency sensitivity will significantly enhance their ability to detect environmental perturbations \cite{2025zwick,2025zwickecc_pop}. An even more sensitive probe of these dynamics may be provided by wide eccentric binaries in the $\mu$Hz regime, which could potentially be observable by LISA \cite{2022xuan}.

Comparable mass binaries embedded in circumbinary discs or AGN discs are expected to contribute to both the ground-based and space-based GW source population \citep{2010Dai,DOrazioCharisi:2023,D'Orazio:binlite:2024,2024whitehead,rowan2023}. In many waveform studies, the influence of the gas is represented through migration torques or laminar drag prescriptions. Numerical simulations, however, consistently show that realistic gas dynamics is characterized by variability, and that simple prescriptions fail to capture the response of the binary. Magneto-rotational turbulence, nonlinear cavity dynamics, shocks, accretion streams, and stochastic inflows can all generate force fluctuations across a broad range of frequencies, and spectral analyses of hydrodynamical simulations \citep[e.g.][]{2005nelson,Roedig_Trqs+2012,2024zwick} show substantial power at various multiples of the orbital frequency. Particularly in thinner discs, where torques become stronger and more coherent, the binary response can be dominated by resonant coupling to these fluctuating components rather than from the mean torque \cite{tiede2020,2024zwick}.

Out of all GW source, EMRIs have the most potential for scientific return. EMRI formation channels are often divided into ``dry'' and ``wet'' scenarios, depending on whether the compact object is delivered to the central black hole through stellar-dynamical processes \cite{Alexander:2003dp} or through interactions with an accretion disk \cite{2021wet}. Estimates suggest that a substantial fraction of observable EMRIs may form in AGN discs \cite{2021wet,2023derdzinski,2026arXiv260529305X}. Additionally, the recently discovered class of quasi-periodic-eruptions \cite{2019Natur.573..381M,2021Natur.592..704A} has a compelling interpretation as a precursor stage to gas embedded EMRIs \cite{2023Linial,2023franchini}, further indicating the abundance of such objects. In gas-embedded EMRIs, the secondary spends hundreds of thousands of orbital cycles immersed in an environment characterized by turbulence, nonlinear hydrodynamics, magnetic fields, shocks, and stochastic accretion flows. At the same time, EMRIs possess several intrinsic frequencies associated with orbital motion and relativistic precession. Environmental variability therefore interacts with a system that is already rich in frequencies, making long-lived resonant coupling expected, and there is already evidence that including eccentricity greatly enhances the possibility of detecting drag effects \cite{2024duque}. 

Similar considerations apply in a variety of other applications. Environmental effects associated with dark matter, ultralight boson fields, or gravitational atoms are frequently represented through effective drag forces~\cite{2022cole,2022coogan,2024PhRvD.109d3504B,2025vicente} on circular orbits. Only in recent work consistently accounting for the backreaction of a relativistic EMRI on a boson cloud environment has the presence of multiple resonances been shown to significantly alter the orbital evolution of eccentric EMRIs~\cite{2026arXiv260503756X, 2026Xu}, as compared to the circular setting~\cite{2023Brito,2025dyson,2025Li,2026Xu,2026Keijzer}. Even in modified-gravity scenarios, additional fluxes or effective forces may possess intrinsic time dependence or stochasticity that cannot be reduced to a simple secular correction~\cite{2026PhRvL.136a1401S,Seymour:2026bjg}.

\subsection{Relativistic extensions}
\label{sec:implications}
In this work we derived the LFH framework using a parametrization of the orbit as a rigidly precessing conic section. Although this captures the fundamental relativistic frequencies that govern resonant coupling to the environment, it is not a complete description of relativistic orbits. For the purposes of this work, the choice of parametrization was a compromise between several competing requirements in view of broad applications. First, sufficient accuracy to actually model smoking gun environmental effects in the context of GW parameter inference. Second, computational efficiency, such that the framework can be used alongside waveform generation packages without significantly increasing their cost. Finally, ease of use for applications in which the perturbation time series is given as a result of numerical simulations, especially for binaries in gas. At present, most such simulations are performed within Newtonian or post-Newtonian frameworks, typically on fixed orbits \cite{Derdzinksi:2021,2021doraziodisc,2023Tiede,Dittmann:Decoupling:2023,2025dittman}. Only recently have some calculations begun to incorporate self-force methods for fully relativistic prescriptions, though treating the disk either as a pressureless medium \cite{2025HegadeKR,2025duque} or as a linearised perfect-fluid \cite{2026Dyson}. Our parametrization based on classical orbital elements together with orbit-averaged precession rates provided an initial compromise between all these aspects. Here, we discuss its limits as well as strategies to extend it to more accurate relativistic schemes.

Firstly, recall that the orbital parametrization is used only within individual adiabatic windows to compute the binary's response to perturbations, while the secular evolution of the carrier orbit itself can be supplied at arbitrarily high accuracy. Within each adiabatic window, the validity of our parameterization can therefore be assessed by comparison with the standard quasi-Keplerian (QK) description of relativistic binaries in the absence of radiation reaction. At 1PN order, the QK parametrization is in fact exactly a rigidly precessing conic section \cite{2004memmesheimer}:
\begin{align}
r &= a \left(1-e_r\cos u\right),\\
M &= u-e_t\sin u,\\
\phi-\phi_0 &= (1+k)v,
\end{align}
where $e_r$, $e_t$, and $e_\phi$ are the radial, temporal, and angular eccentricities, $\phi$ and $v$ are orbital phases and $k$ is the apsidal precession parameter. Relativistic effects are captured through the distinction between these eccentricities, which is not modeled in the present work, and through the precession rate $k$, which is instead directly equivalent to our choice of $\dot\omega$. Importantly however, these quasi-Keplerian relations introduce no additional frequencies to the dynamics. Rather, they modify the mapping between orbital energy and angular momentum and the osculating geometry of the orbit. As a result, the resonant response developed in this work carries over directly once the appropriate quasi-Keplerian parameters are employed and the LPE are extended.

The situation changes at higher post-Newtonian orders. Beginning at 2PN order for non-spinning binaries, the quasi-Keplerian parametrization acquires additional oscillatory contributions through terms of the form \cite{2004memmesheimer}:
\begin{align}
M
&=
u-e_t\sin u
+
f_t\sin v
+
g_t(v-u),
\\
\phi-\phi_0
&=
(1+k)v
+
f_\phi\sin 2v
+
g_\phi\sin 3v,
\end{align}
where the new coefficients $f_t$, $f_{\phi}$ and $g_{\phi}$ depend on the orbital energy and angular momentum. These corrections generate harmonics beyond those present in the rigidly precessing ellipse approximation. In principle, such harmonics can participate in resonant interactions with the environment and therefore produce signatures that are absent from the present treatment. Whether these higher order corrections are observable depends on both the strength of the resonance and the signal-to-noise ratio of the source. In practice, most astrophysical environmental effects are dynamically important primarily during the early inspiral, where a leading-order orbital parametrization is generally sufficient for computing the additional perturbations. Nevertheless, an important extension of the present framework is to replace the rigidly precessing ellipse by a higher-order quasi-Keplerian carrier. This would allow the LFH formalism to be applied to arbitrary perturbations. Work in this direction is currently underway.

An important complementary regime where the above argument does not apply cleanly is in the case of EMRIs. EMRIs are modelled through self-force theory, a means of expanding the orbital dynamics in powers of the mass ratio. Unlike other perturbative approaches, self-force theory remains valid all the way to merger, providing rigorous control over the fully relativistic dynamics of small bodies near supermassive black holes. The framework presented in this work can be extended to this regime through 
the method of osculating geodesics. In vacuum around a spinning black hole, a generic EMRI trajectory ergodically fills a torus surrounding the central object from which notions of the semi-latus rectum, eccentricity, and inclination can be defined for fully relativistic orbits. Parameterizing the orbit through Mino time $\lambda$, which extends the separable symmetries of spinning black holes, one finds that generic orbits are described by two fundamental {\it Mino frequencies} $\Upsilon_r$ and $\Upsilon_\theta$, such that the orbital phases evolve as $q_r = \Upsilon_r\lambda + q_r^0$ and $q_\theta = \Upsilon_\theta\lambda + q_\theta^0$, generalising the notion of mean anomaly to orbits around spinning black holes. Collecting the orbital constants $P_i = \{p, e, I\}$ and phases $q_i = \{q_r, q_\theta\}$, and introducing a small perturbing force, one obtains the perturbed osculating geodesic equations,
\begin{align}
    \dot{P}_i &= \epsilon F_i(P,q) + \cdots\\
    \dot{q}_i &= \Upsilon_i(P) + \epsilon f_i(P,q) + \cdots
\end{align}
as described in \cite{2024Lynch}, acting as a direct extension of the LPE in the fully relativistic but small mass-ratio limit.

Orbital resonances of EMRIs around spinning black holes have received significant attention, both in the vacuum case where self-resonances occur when $\Upsilon_r/\Upsilon_\theta \in \mathbb{Q}$ \cite{2015Brink, 2012Flanagan, 2013vandeMeent, 2024Lynch}, and in the presence of external tidal fields  \cite{2019Yang, 2019Bonga, 2021Gupta, 2022Gupta}, in analogy with the Newtonian context discussed above. However, attention to  environments with generic structure is largely absent from the relativistic EMRI  literature. Work  is however currently underway by some of us in connecting forces on the EMRI dynamics to locally integrated torques exerted on the environment \cite{2026Dyson, 2026Dyson3rd}, which we expect will provide a direct route towards calculating such forcing functions in a tractable manner. A self-force extension of the LFH formalism, combined with these relativistic torque balance relations and the multiscale  expansions standard in self-force theory \cite{2025Lewis, 2025Mathews}, would  offer an efficient means to model the co-evolution of fully relativistic EMRIs with realistic astrophysical environments.

\section{Conclusion}
\label{sec:conclusion}
The next generation of GW observatories will detect more and higher quality signals originating from a wide range of source. Third-generation ground-based detectors such as ET and CE will probe populations of eccentric binaries from the dynamical and AGN channels. These sources are prime candidates to exhibit environmental effects, and their detection would allow to pin-point their origin and characterize their immediate surroundings. Space based observatories such as LISA and TianQin \cite{TianQin} will observe a diverse sources systems ranging from massive black-hole binaries to EMRIs. Such systems typically form in the dense environments of galactic nuclei, and the detection of environmental effects in their gravitational radiation offers a new way to probe matter in the vicinity of massive BHs. However, extracting the maximum amount of science from these observations require models of environmental effects at level of rigor comparable to that already achieved for vacuum dynamics. In this work, we have developed the Lagrange--Fourier--Hansen framework as a tool to provide such models. The central idea is to combine the Lagrange Planetary Equations, Fourier spectral decompositions, and expansions in Hansen coefficients into a method that can efficiently connect arbitrary force time series to the orbital evolution evolution of relativistic binaries. A key motivation for this work is that realistic astrophysical environments are characterized by time-variable and even stochastic forcings. Gas turbulence, nonlinear hydrodynamics, external tides, stellar encounters, dark-matter substructure, and many other processes produce forcing characterized by time variability and stochasticity. Eccentric and precessing binaries respond to perturbations that remain phase coherent with the orbital motion, which can produce rich dynamical behaviour and informative GW signatures. The goal of LFH framework is to provide a systematic way of identifying these components, quantify their contribution to the GW signal, and reveal smoking gun observables of different binary environments. Several extensions to the derivations presented in this work are already underway. On the practical side, the LFH formalism is being incorporated into the forthcoming \texttt{AstroWaveforms} package. On the theoretical side, the most immediate extension is the replacement of the rigidly precessing ellipse by higher-order quasi-Keplerian and osculating geodesic parametrization.

\section*{Data availability}
The code used to produce the results will be shared upon reasonable request.
\section*{Acknowledgments}
L.Z. is supported by the European Union’s Horizon 2024 research and innovation program under the Marie Sklodowska-Curie grant agreement No. 101208914. The Center of Gravity is a Center of Excellence funded by the Danish National Research Foundation under grant No. 184. CD acknowledges support by VILLUM Foundation (grant no.VIL37766 and no. VIL53101) and the DNRF Chair program (grant no. DNRF162) by the Danish National Research Foundation. JS is supported by the Villum Fonden grant No. 29466,
and by the ERC Starting Grant no. 101043143 – Black-
HoleMergs. LLMs were used for formatting, coding and for ideation.

\appendix
\onecolumngrid
\section{Resonant Response for Rigidly Precessing Elliptical Orbits with Radiation Reaction.}
\label{Appendix:Resonant_response}
\subsection{Adiabatic window, precessing carrier orbit, and Lagrange planetary equations}
\label{Appendix:Resonant_response:adiabatic}

Our first task is to identify the response of a relativistic binary to weak environmental perturbations. We work within a local adiabatic time window:
\begin{align}
\mathcal I_\tau
\equiv
\left[
\tau-\frac{\Delta t_W}{2},
\,\,
\tau+\frac{\Delta t_W}{2}
\right],
\qquad
T_{\rm orb}\ll \Delta t_W \ll T_{\rm rr},
\label{eq:window}
\end{align}
where $T_{\rm orb}$ is a characteristic orbital timescale and $T_{\rm rr}$ is the radiation-reaction timescale. The parameter $\tau$ denotes the central epoch of the window. Within $\mathcal I_\tau$, the binary completes many orbital cycles while the radiation reaction driven inspiral remains negligible. Inside each adiabatic window, we approximate the vacuum binary as a conservative relativistic orbit with slowly varying precessional frequencies. The exact parametrization is given in terms of classic orbital elements:
\begin{align}
\mathbf Y_{\rm c}
=
\bigl(
p_{\rm c},
e_{\rm c},
\omega_{\rm c},
\Omega_{\rm c},
\iota_{\rm c},M_{\rm c}
\bigr)^T,
\end{align}
where $p_{\rm c}$ and $e_{\rm c}$ are the semi-latus rectum and eccentricity, while $(\omega_{\rm c},\Omega_{\rm c},\iota_{\rm c})$ specify the orientation of the precessing orbit. We further introduce the carrier mean anomaly $M_{\rm c}$ together with
the mean motion $ \bar n_{\rm c}$. The true anomaly $\nu_{\rm c}$ parametrizes the radial motion:
\begin{align}
r(\nu_{\rm c})
=
\frac{p_{\rm c}}{1+e_{\rm c}\cos\nu_{\rm c}},
\label{eq:carrier_radius}
\end{align}
with $R$ denoting the binary separation. Within the local window, the angles may be linearized as:
\begin{align}
\omega_{\rm c}(t)
&\simeq
\omega_{\rm c}(\tau)
+
\dot\omega_{\rm c}(\tau)(t-\tau),
\\
\Omega_{\rm c}(t)
&\simeq
\Omega_{\rm c}(\tau)
+
\dot\Omega_{\rm c}(\tau)(t-\tau),
\\
\iota_{\rm c}(t)
&\simeq
\iota_{\rm c}(\tau)
+
\dot\iota_{\rm c}(\tau)(t-\tau),\\
M_{\rm c}(t)
&\simeq
M_{\rm c}(\tau)
+
\bar n_{\rm c}(\tau)(t-\tau),
\label{eq:carrier_precession}
\end{align}
where the precession frequencies and mean motion vary only on the slow inspiral timescale. Consequently, the orbital geometry and frequencies may be treated as constant within $\mathcal I_\tau$. The carrier orbit is embedded in Cartesian space through the Euler rotations:
\begin{align}
\mathbf r(\nu_{\rm c})
=
\mathcal R_z(\Omega_{\rm c})
\mathcal R_x(\iota_{\rm c})
\mathcal R_z(\omega_{\rm c})
\begin{pmatrix}
R\cos\nu_{\rm c}\\
R\sin\nu_{\rm c}\\
0
\end{pmatrix},
\label{eq:carrier_embedding}
\end{align}
where $\mathcal R_i$ denotes a rotation about the $i$-axis. We now introduce a weak environmental perturbation. The vectors $\mathbf a_1$ and $\mathbf a_2$ denote the non-vacuum accelerations acting on the two binary components. The perturbing relative acceleration is:
\begin{align}
\mathbf a_{\rm env}
\equiv
\mathbf a_1-\mathbf a_2.
\end{align}
We decompose $\mathbf a_{\rm env}$ into the orbital basis:
\begin{align}
\mathbf a_{\rm env}
=
R\,\hat{\mathbf r}
+
S\,\hat{\bm\phi}
+
W\,\hat{\mathbf w},
\label{eq:force_decomposition}
\end{align}
where $R$, $S$, and $W$ are the radial, tangential, and normal force components. The orbital response is described through variation of constants applied to the precessing carrier orbit. We write the total orbital elements as:
\begin{align}
\mathbf Y
=
\mathbf Y_{\rm c}
+
\mathbf Y_{\rm p},
\end{align}
where $\mathbf Y_{\rm p}$ denotes the environmental perturbation. At any given moment, our carrier orbit differs from a fully Newtonian orbit simply by a slow rigid rotation. To first order in the environmental forcing, the carrier orbit may be regarded as a prescribed background solution. The perturbation enters only through the additional acceleration projected onto the instantaneous orbital triad, while the rigid precession of the carrier is absorbed into the unperturbed evolution. Consequently, the variation-of-constants derivation proceeds unchanged and yields the standard Gauss form of the LPE:
\begin{align}
\dot p_{\rm p}
&=
2\sqrt{\frac{p_{\rm c}}{Gm}}\,r
\,
S_{\rm c},
\label{eq:lpe_p}
\\
\dot e_{\rm p}
&=
\sqrt{\frac{p_{\rm c}}{Gm}}
\left[
\sin\nu_{\rm c}\,R_{\rm c}
+
\left(
\cos\nu_{\rm c}
+
\frac{e_{\rm c}+\cos\nu_{\rm c}}
{1+e_{\rm c}\cos\nu_{\rm c}}
\right)S_{\rm c}
\right],
\label{eq:lpe_e}
\\
\dot\iota_{\rm p}
&=
\sqrt{\frac{p_{\rm c}}{Gm}}\,
\frac{r}{p_{\rm c}}\,
\cos(\omega_{\rm c}+\nu_{\rm c})\,W_{\rm c},
\label{eq:lpe_i}
\\
\dot\Omega_{\rm p}
&=
\sqrt{\frac{p_{\rm c}}{Gm}}\,
\frac{r}{p_{\rm c}}\,
\frac{\sin(\omega_{\rm c}+\nu_{\rm c})}
{\sin\iota_{\rm c}}\,
W_{\rm c},
\label{eq:lpe_Omega}
\\
\dot\omega_{\rm p}
&=
\sqrt{\frac{p_{\rm c}}{Gm}}
\Biggl[
-\frac{\cos\nu_{\rm c}}{e_{\rm c}}\,R_{\rm c}
+
\frac{\sin\nu_{\rm c}}{e_{\rm c}}
\left(
1+\frac{r}{p_{\rm c}}
\right)S_{\rm c}
\nonumber\\
&\qquad
-
\frac{r}{p_{\rm c}}\,
\cot\iota_{\rm c}\,
\sin(\omega_{\rm c}+\nu_{\rm c})\,W_{\rm c}
\Biggr],
\label{eq:lpe_omega}
\end{align}
where the equation for the mean anomaly is treated separately in Appendix \ref{Appendix:Resonant_response:Mean_anomaly}.

\subsection{Fourier--Hansen decomposition of the forcing problem}
\label{Appendix:Resonant_response:Hansen}

Working within an adiabatic window we introduce a local Fourier decomposition of the projected environmental force:
\begin{align}
\mathbf a_{\rm env}(t)
\approx
\frac{1}{2\pi}
\int_{-\infty}^{\infty}
\tilde {\mathcal{W}}_{\mathcal I}(\sigma)\,
\tilde{\mathbf a}_{\rm env}(\tau;\sigma)\,
e^{i\sigma t}
\,d\sigma,
\label{eq:local_force_decomposition}
\end{align}
with local spectral amplitudes:
\begin{equation}
\tilde{\mathbf a}_{\rm env}(\tau;\sigma)
\equiv\int dt' \mathcal W_\mathcal{I}(t')\mathbf a_{\rm env}(t')\,
e^{-i\sigma t'}
\,dt',
\end{equation}
where $\mathcal W_{\mathcal I}$ denotes an appropriate windowing kernel that corresponds to the finite integration over the interval $\mathcal I$ and:
\begin{align}
    \Delta t_{W}\equiv\int_{-\infty}^{\infty}  \mathcal W_{\mathcal I}(t) \,dt .
\end{align}
We now expand the orbital dependence of the Lagrange planetary equations using Hansen coefficients. Introducing the semi-major axis:
\begin{align}
a_{\rm c}
\equiv
\frac{p_{\rm c}}{1-e_{\rm c}^2},
\end{align}
the standard Hansen expansion reads:
\begin{align}
\left(
\frac{r}{a_{\rm c}}
\right)^j
e^{ik\nu_{\rm c}}
=
\sum_{l=-\infty}^{\infty}
X_l^{jk}(e_{\rm c})\,
e^{ilM_{\rm c}}.
\label{eq:hansen_expansion}
\end{align}

We further define the cosine and sine combinations:
\begin{align}
\mathcal C_l^{jk}(e_{\rm c})
&\equiv
\frac{1}{2}
\left[
X_l^{j,k}(e_{\rm c})
+
X_l^{j,-k}(e_{\rm c})
\right],
\\
\mathcal S_l^{jk}(e_{\rm c})
&\equiv
\frac{1}{2i}
\left[
X_l^{j,k}(e_{\rm c})
-
X_l^{j,-k}(e_{\rm c})
\right].
\label{eq:hansen_CS}
\end{align}

The first-order LPE then become:
\begin{align}
\dot p_{\rm p}
&=
2a_{\rm c}
\sqrt{\frac{p_{\rm c}}{Gm}}
\sum_l
X_l^{1,0}\,
e^{ilM_{\rm c}}\,
S,
\label{eq:lpe_p_hansen}
\\
\dot e_{\rm p}
&=
\sqrt{\frac{p_{\rm c}}{Gm}}
\sum_l
e^{ilM_{\rm c}}
\Biggl[
\mathcal S_l^{0,1}\,
R
+
\left(
\mathcal C_l^{0,1}
+
e_{\rm c}\frac{a_{\rm c}}{p_{\rm c}}X_l^{1,0}
+
\frac{a_{\rm c}}{p_{\rm c}}
\mathcal C_l^{1,1}
\right)
S
\Biggr],
\label{eq:lpe_e_hansen}
\\
\dot\iota_{\rm p}
&=
\frac{a_{\rm c}}{2p_{\rm c}}
\sqrt{\frac{p_{\rm c}}{Gm}}
\sum_l
\left[
X_l^{1,1}\,
e^{i(lM_{\rm c}+\omega_{\rm c})}
+
X_l^{1,-1}\,
e^{i(lM_{\rm c}-\omega_{\rm c})}
\right]
W,
\label{eq:lpe_i_hansen}
\\
\dot\Omega_{\rm p}
&=
\frac{a_{\rm c}}{2ip_{\rm c}}
\sqrt{\frac{p_{\rm c}}{Gm}}
\frac{1}{\sin\iota_{\rm c}}
\sum_l
\left[
X_l^{1,1}\,
e^{i(lM_{\rm c}+\omega_{\rm c})}
-
X_l^{1,-1}\,
e^{i(lM_{\rm c}-\omega_{\rm c})}
\right]
W,
\label{eq:lpe_Omega_hansen}
\\
\dot\omega_{\rm p}
&=
\sqrt{\frac{p_{\rm c}}{Gm}}
\sum_l
e^{ilM_{\rm c}}
\Biggl[
-
\frac{1}{e_{\rm c}}
\mathcal C_l^{0,1}\,
R
+
\left(
\frac{1}{e_{\rm c}}
\mathcal S_l^{0,1}
+
\frac{1}{e_{\rm c}}
\frac{a_{\rm c}}{p_{\rm c}}
\mathcal S_l^{1,1}
\right)
S
\Biggr]
\nonumber\\
&\qquad
-
\frac{a_{\rm c}}{2ip_{\rm c}}
\sqrt{\frac{p_{\rm c}}{Gm}}
\cot\iota_{\rm c}
\sum_l
\left[
X_l^{1,1}\,
e^{i(lM_{\rm c}+\omega_{\rm c})}
-
X_l^{1,-1}\,
e^{i(lM_{\rm c}-\omega_{\rm c})}
\right]
W.
\label{eq:lpe_omega_hansen}
\end{align}

Substituting the local Fourier decomposition of the force components into the LPE yields:
\begin{align}
\dot p_{\rm p}
&=
2a_{\rm c}
\sqrt{\frac{p_{\rm c}}{Gm}}
\sum_l
X_l^{1,0}
\int
\tilde S(\tau;\sigma)\,
e^{i\Phi_l^{(0)}(\tau;t)}
\,d\sigma,
\label{eq:lpe_p_fourier}
\\
\dot e_{\rm p}
&=
\sqrt{\frac{p_{\rm c}}{Gm}}
\sum_l
\int
\Biggl[
\mathcal S_l^{0,1}\,
\tilde R(\tau;\sigma)
\nonumber\\
&\qquad
+
\left(
\mathcal C_l^{0,1}
+
e_{\rm c}\frac{a_{\rm c}}{p_{\rm c}}X_l^{1,0}
+
\frac{a_{\rm c}}{p_{\rm c}}
\mathcal C_l^{1,1}
\right)
\tilde S(\tau;\sigma)
\Biggr]
e^{i\Phi_l^{(0)}(\tau;t)}
\,d\sigma,
\label{eq:lpe_e_fourier}
\\
\dot\iota_{\rm p}
&=
\frac{a_{\rm c}}{2p_{\rm c}}
\sqrt{\frac{p_{\rm c}}{Gm}}
\sum_l
\int
\Bigl[
X_l^{1,1}\,
e^{i\Phi_l^{(+1)}(\tau;t)}
+
X_l^{1,-1}\,
e^{i\Phi_l^{(-1)}(\tau;t)}
\Bigr]
\tilde W(\tau;\sigma)
\,d\sigma,
\label{eq:lpe_i_fourier}
\\
\dot\Omega_{\rm p}
&=
\frac{a_{\rm c}}{2ip_{\rm c}}
\sqrt{\frac{p_{\rm c}}{Gm}}
\frac{1}{\sin\iota_{\rm c}}
\sum_l
\int
\Bigl[
X_l^{1,1}\,
e^{i\Phi_l^{(+1)}(\tau;t)}
-
X_l^{1,-1}\,
e^{i\Phi_l^{(-1)}(\tau;t)}
\Bigr]
\tilde W(\tau;\sigma)
\,d\sigma,
\label{eq:lpe_Omega_fourier}
\\
\dot\omega_{\rm p}
&=
\sqrt{\frac{p_{\rm c}}{Gm}}
\sum_l
\int
\Biggl[
-
\frac{1}{e_{\rm c}}
\mathcal C_l^{0,1}\,
\tilde R(\tau;\sigma)
+
\left(
\frac{1}{e_{\rm c}}
\mathcal S_l^{0,1}
+
\frac{1}{e_{\rm c}}
\frac{a_{\rm c}}{p_{\rm c}}
\mathcal S_l^{1,1}
\right)
\tilde S(\tau;\sigma)
\Biggr]
e^{i\Phi_l^{(0)}(\tau;t)}
\,d\sigma
\nonumber\\
&\qquad
-
\frac{a_{\rm c}}{2ip_{\rm c}}
\sqrt{\frac{p_{\rm c}}{Gm}}
\cot\iota_{\rm c}
\sum_l
\int
\Bigl[
X_l^{1,1}\,
e^{i\Phi_l^{(+1)}(\tau;t)}
-
X_l^{1,-1}\,
e^{i\Phi_l^{(-1)}(\tau;t)}
\Bigr]
\tilde W(\tau;\sigma)
\,d\sigma,
\label{eq:lpe_omega_fourier}
\end{align}
where we introduced the total phases:
\begin{align}
\Phi_l^{(q)}(\tau;t)
\equiv
\sigma t
+
lM_{\rm c}
+
q\omega_{\rm c},
\qquad
q=0,\pm1.
\label{eq:total_phase}
\end{align}
Here $q=0$ corresponds to the orbital harmonics and $q=\pm1$ to the apsidal sidebands generated by the factors $e^{\pm i\omega_{\rm c}}$ appearing in the LPE. Within the local window, the phases may be linearized as:
\begin{align}
\Phi_l^{(q)}(\tau;t)
\simeq
\Phi_l^{(q)}(\tau)
+
\Delta_l^{(q)}(\sigma;\tau)
(t-\tau),
\label{eq:phase_linearization}
\end{align}
with local detuning frequencies:
\begin{align}
\Delta_l^{(q)}(\sigma;\tau)
=
\sigma
+
l\bar n_{\rm c}
+
q\dot\omega_{\rm c}.
\label{eq:detuning}
\end{align}
Resonant forcing with the orbital epi-cycles can occur whenever the local spectrum of the projected force develops support near:
\begin{align}
\sigma
\simeq
-
l \bar n_{\rm c}
-
q\dot\omega_{\rm c}.
\label{eq:resonance_condition}
\end{align}
In addition to the harmonic resonances above, the projected force may also contain slowly varying secular components concentrated near $\sigma\simeq0$. These contributions are encoded directly in the local spectrum $\tilde{\mathbf a}_{\rm env}(\tau;\sigma)$ and evolve through their explicit dependence on the window center $\tau$.

\subsection{Selection of the resonant modes via window averaging}
\label{Appendix:Resonant_response:selection}
The results above are expressed as sums over $l$-harmonics and integrals over Fourier modes in $\sigma$. For any orbital element $x_{\rm p}$ or its time derivative $\dot x_{\rm p}$, we define the local average over the adiabatic interval $\mathcal I_\tau$ as:
\begin{align}
\langle \dot x_{\rm p} \rangle_{\mathcal I}(\tau)
\equiv
\frac{1}{\Delta t_W}
\int_{\mathcal I_\tau}
\dot x_{\rm p}(t)\,dt.
\label{eq:local_secular_average}
\end{align}
Following the phase linearization of Eq.~\ref{eq:phase_linearization}, we substitute into the averaging integral:
\begin{align}
\langle \dot x_{\rm p} \rangle_{\mathcal I}
=\frac{1}{\Delta t_W}\sum_l
\int_{\mathcal I_\tau}\int \mathcal W_{\mathcal I}(\sigma)
 H_{\chi}^{(q)}\tilde A(\tau;\sigma)\,
e^{i\Phi_l^{(q)}(\tau;t)}
\,d\sigma\,dt,
\end{align}
where $A\in(R,S,W)$ is the appropriate acceleration and $H_l^{(q)}$ is the appropriate Hansen coefficient for the LPE of each orbital element.
Evaluating the integral over the finite window yields:
\begin{align}
\langle \dot x_{\rm p} \rangle_{\mathcal I}
&\simeq \sum_l
 \tilde A^{(q)}(\tau;\sigma)\,
e^{i\Phi_l^{(q)}(\tau)}
\operatorname{sinc}\!\left(
\frac{
\Delta_l^{(q)}(\sigma;\tau)\Delta t_W
}{2}
\right),
\label{eq:sinc_window_filter}
\end{align}
where $\Phi_l^{(q)}(\tau)$ denotes the total phase evaluated at the window center. The sinc kernel acts as a local spectral filter, selecting harmonics satisfying:
\begin{align}
|\Delta_l^{(q)}|
\lesssim
\Delta t_W^{-1}.
\end{align}
In the long-window limit:
\begin{align}
\operatorname{sinc}\!\left(
\frac{
\Delta_l^{(q)}(\sigma;\tau)\Delta t_W
}{2}
\right)
\longrightarrow
\frac{2\pi}{\Delta t_W}
\delta\!\left(
\Delta_l^{(q)}(\sigma;\tau)
\right),
\label{eq:local_delta_limit}
\end{align}
which selects the resonant frequencies:
\begin{align}
\sigma
=
-\sigma_l^{(q)},
\qquad
\sigma_l^{(q)}
\equiv
l\bar n_{\rm c}
+
q\dot\omega_{\rm c}.
\label{eq:local_resonant_frequency}
\end{align}
We define the residual phase:
\begin{align}
\psi_l^{(q)}(\tau)
&\equiv
\Phi_l^{(q)}(\tau;t)
\bigg|_{\sigma=-\sigma_l^{(q)}}=
-\sigma_l^{(q)}\tau
+
lM_{\rm c}(\tau)
+
q\omega_{\rm c}(\tau).
\label{eq:resonant_phase_psi}
\end{align}
This phase is slowly varying over a radiation reaction timescale. It keeps track of the phase alignment between the local Fourier coefficient and the corresponding Hansen channel. Substituting the long-window limit into the Fourier--Hansen expanded LPE yields the leading adiabatic secular evolution:
\begin{align}
\langle \dot p_{\rm p} \rangle_{\mathcal I}^{\rm res}
&\simeq
\frac{2a_{\rm c}}{\Delta t_W}
\sqrt{\frac{p_{\rm c}}{Gm}}
\sum_l
X_l^{1,0}\,
\tilde S\!\left(
-\sigma_l^{(0)};\tau
\right)
e^{i\psi_l^{(0)}(\tau)},
\label{eq:pdot_secular_selected}
\\[1ex]
\langle \dot e_{\rm p} \rangle_{\mathcal I}^{\rm res}
&\simeq
\frac{1}{\Delta t_W}
\sqrt{\frac{p_{\rm c}}{Gm}}
\sum_l
\Biggl[
\mathcal S_l^{0,1}\,
\tilde R\!\left(
-\sigma_l^{(0)};\tau
\right)
+
\left(
\mathcal C_l^{0,1}
+
e_{\rm c}\frac{a_{\rm c}}{p_{\rm c}}X_l^{1,0}
+
\frac{a_{\rm c}}{p_{\rm c}}\mathcal C_l^{1,1}
\right)
\tilde S\!\left(
-\sigma_l^{(0)};\tau
\right)
\Biggr]
e^{i\psi_l^{(0)}(\tau)},
\label{eq:edot_secular_selected}
\\[1ex]
\langle \dot\iota_{\rm p} \rangle_{\mathcal I}^{\rm res}
&\simeq
\frac{a_{\rm c}}{2p_{\rm c}\Delta t_W}
\sqrt{\frac{p_{\rm c}}{Gm}}
\sum_l
\Biggl[
X_l^{1,1}\,
\tilde W\!\left(
-\sigma_l^{(+1)};\tau
\right)
e^{i\psi_l^{(+1)}(\tau)}
+
X_l^{1,-1}\,
\tilde W\!\left(
-\sigma_l^{(-1)};\tau
\right)
e^{i\psi_l^{(-1)}(\tau)}
\Biggr],
\label{eq:iotadot_secular_selected}
\\[1ex]
\langle \dot\Omega_{\rm p} \rangle_{\mathcal I}^{\rm res}
&\simeq
\frac{a_{\rm c}}{2ip_{\rm c}\Delta t_W\sin\iota_{\rm c}}
\sqrt{\frac{p_{\rm c}}{Gm}}
\sum_l
\Biggl[
X_l^{1,1}\,
\tilde W\!\left(
-\sigma_l^{(+1)};\tau
\right)
e^{i\psi_l^{(+1)}(\tau)}
-
X_l^{1,-1}\,
\tilde W\!\left(
-\sigma_l^{(-1)};\tau
\right)
e^{i\psi_l^{(-1)}(\tau)}
\Biggr],
\label{eq:Omegadot_secular_selected}
\\[1ex]
\langle \dot\omega_{\rm p} \rangle_{\mathcal I}^{\rm res}
&\simeq
\frac{1}{\Delta t_W}
\sqrt{\frac{p_{\rm c}}{Gm}}
\sum_l
\Biggl[
-
\frac{1}{e_{\rm c}}
\mathcal C_l^{0,1}\,
\tilde R\!\left(
-\sigma_l^{(0)};\tau
\right)
\nonumber\\
&\qquad
+
\frac{1}{e_{\rm c}}
\left(
\mathcal S_l^{0,1}
+
\frac{a_{\rm c}}{p_{\rm c}}
\mathcal S_l^{1,1}
\right)
\tilde S\!\left(
-\sigma_l^{(0)};\tau
\right)
\Biggr]
e^{i\psi_l^{(0)}(\tau)}
-
\frac{\cot\iota_{\rm c}}{\sin\iota_{\rm c}}
\langle \dot\Omega_{\rm p} \rangle_{\mathcal I}^{\rm res}
\sin\iota_{\rm c}.
\label{eq:omegadot_secular_selected}
\end{align}
The factors:
\begin{align}
\tilde R\!\left(-\sigma_l^{(q)};\tau\right)e^{i\psi_l^{(q)}(\tau)},
\qquad
\tilde S\!\left(-\sigma_l^{(q)};\tau\right)e^{i\psi_l^{(q)}(\tau)},
\qquad
\tilde W\!\left(-\sigma_l^{(q)};\tau\right)e^{i\psi_l^{(q)}(\tau)}
\end{align}
have an amplitude and a phase. Their phases determine whether consecutive windows add coherently or destructively. The secular response is therefore determined jointly by the local spectral support of the projected environmental forcing, the Hansen couplings of the LPE, and the phase alignment between the forcing and the carrier orbit.

\subsection{Response of the binary's mean anomaly}
\label{Appendix:Resonant_response:Mean_anomaly}
We now close the system by deriving the perturbation of the intrinsic fast orbital phase, which we take to be the mean anomaly $M$. The LPE for the mean anomaly is:
\begin{align}
\dot M
=
\bar n
-
\frac{2r}{\bar n a^2}R
+
\sqrt{1-e^2}
\left(
\dot\omega
+
\cos\iota\,\dot\Omega
\right),
\label{eq:mean_anomaly_lpe}
\end{align}
The first term describes the Keplerian sweep along the osculating orbit, the second term is a direct radial forcing contribution, while the remaining terms correspond to geometric re-orientations of the orbit. We linearize Eq.~\ref{eq:mean_anomaly_lpe} about the carrier orbit. Varying the mean motion gives:
\begin{align}
\bar n_{\rm p}
=
-\frac{3}{2}
\bar n_{\rm c}
\frac{a_{\rm p}}{a_{\rm c}}=-\frac{3}{2}
\bar n_{\rm c}
\frac{p_{\rm p}}{p_{\rm c}}
-
\frac{3e_{\rm c}}{1-e_{\rm c}^2}
\bar n_{\rm c}
e_{\rm p},
\label{eq:mean_motion_variation}
\end{align}
The perturbation equation for the mean anomaly therefore becomes:
\begin{align}
\dot M_{\rm p}
=
f_M^{\rm Kep}
+
f_M^{\rm dir}
+
f_M^{\rm geom},
\label{eq:mean_anomaly_response}
\end{align}
with:
\begin{align}
f_M^{\rm Kep}
\equiv
-
\frac{3}{2}
\bar n_{\rm c}
\frac{a_{\rm p}}{a_{\rm c}},\qquad
f_M^{\rm dir}
\equiv
-
\frac{2r}{\bar n_{\rm c}a_{\rm c}^2}
R,\qquad
f_M^{\rm geom}
\equiv
\sqrt{1-e_{\rm c}^2}
\left(
\dot\omega_{\rm p}
+
\cos\iota_{\rm c}\dot\Omega_{\rm p}
\right).
\label{eq:mean_anomaly_force_split}
\end{align}
Reintroducing the adiabatic window $\mathcal I_\tau$, we define the averaged mean-anomaly drift by averaging Eq.~\ref{eq:mean_anomaly_response} over the local window:
\begin{align}
\langle
\dot M_{\rm p}
\rangle_{\mathcal I}
=
\langle
\dot M_{\rm p}
\rangle_{\mathcal I}^{\rm Kep}
+
\langle
\dot M_{\rm p}
\rangle_{\mathcal I}^{\rm dir}
+
\langle
\dot M_{\rm p}
\rangle_{\mathcal I}^{\rm geom}.
\label{eq:mean_anomaly_avg_split}
\end{align}

\subsubsection{Secular contributions}
\label{Appendix:Resonant_response:VI_b}

The secular part of Keplerian contribution follows directly from the variation of the osculating mean motion and the growth of $a_{\rm p}$:
\begin{align}
\langle
\dot M_{\rm p}
\rangle_{\mathcal I}^{\rm Kep}
=
-
\frac{3}{2}
\bar n_{\rm c}
\frac{
a_{\rm p}(\tau)
}{
a_{\rm c}
}.
\label{eq:mean_history_avg}
\end{align}
This term describes the secular accumulation of orbital phase generated by the historical drift of the binary mean motion. The contribution from direct radial forcing is:
\begin{align}
\langle
\dot M_{\rm p}
\rangle_{\mathcal I}^{\rm dir}
=
\langle-
\frac{2R}{\bar n_{\rm c}a_{\rm c}^2}
R\rangle_{\mathcal I}.
\label{eq:mean_direct_sector}
\end{align}
We use the same Fourier-Hansen expansion and obtain the secular growth:
\begin{align}
\langle
\dot M_{\rm p}
\rangle_{\mathcal I}^{\rm dir}
=
-
\frac{2}{\Delta t_W}
\frac{1}{\bar n_{\rm c}a_{\rm c}}
\sum_l
X_l^{1,0}
\,
\tilde R
\left(
\tau;
-\sigma_l^{(0)}
\right)
e^{i\psi_l^{(0)}(\tau)}.
\label{eq:mean_direct_avg}
\end{align}
This contribution represents the direct modification of the orbital sweep rate by resonant radial forcing evaluated along the carrier orbit. The remaining contribution corresponds to geometric reorientations of the orbit:
\begin{align}
\langle
\dot M_{\rm p}
\rangle_{\mathcal I}^{\rm geom}
=
\sqrt{1-e_{\rm c}^2}
\left(
\langle
\dot\omega_{\rm p}
\rangle_{\mathcal I}
+
\cos\iota_{\rm c}
\langle
\dot\Omega_{\rm p}
\rangle_{\mathcal I}
\right).
\label{eq:mean_geometric_avg}
\end{align}
This term represents the change in the inertial orientation of the osculating ellipse relative to the carrier orbit. The averaged mean-anomaly response therefore becomes:
\begin{align}
\langle
\dot M_{\rm p}
\rangle_{\mathcal I}
&=
-
\frac{3}{2}
\bar n_{\rm c}
\frac{
a_{\rm p}(\tau)
}{
a_{\rm c}
}
-
\frac{2}{\Delta t_W}
\frac{1}{\bar n_{\rm c}a_{\rm c}}
\sum_l
X_l^{1,0}
\,
\tilde R
\left(
\tau;
-\sigma_l^{(0)}
\right)
e^{i\psi_l^{(0)}(\tau)}
+
\sqrt{1-e_{\rm c}^2}
\left(
\langle
\dot\omega_{\rm p}
\rangle_{\mathcal I}
+
\cos\iota_{\rm c}
\langle
\dot\Omega_{\rm p}
\rangle_{\mathcal I}
\right).
\label{eq:mean_anomaly_final_avg}
\end{align}

\subsection{Integration over radiation reaction timescales}
\label{Appendix:Resonant_response:radiation_reaction}
\subsubsection{The carrier over radiation reaction timescales}

Over the radiation-reaction timescale $T_{\rm rr}$, the assumption of a conservative carrier orbit fails. Instead, we treat the unperturbed carrier as a dynamical reference trajectory that shrinks and circularizes according to orbit-averaged radiation reaction. The carrier elements evolve as:
\begin{align}
p_{\rm c}(\tau)
&=
p_{{\rm c},0}
+
\int_{\tau_0}^{\tau}
\dot p_{\rm c}
\bigl(
p_{\rm c}(\tau'),e_{\rm c}(\tau')
\bigr)
\,d\tau',
\label{eq:carrier_flow_p}
\\
e_{\rm c}(\tau)
&=
e_{{\rm c},0}
+
\int_{\tau_0}^{\tau}
\dot e_{\rm c}
\bigl(
p_{\rm c}(\tau'),e_{\rm c}(\tau')
\bigr)
\,d\tau'.
\label{eq:carrier_flow_e}
\end{align}
The carrier phases are evolved explicitly as:
\begin{align}
M_{\rm c}(\tau)
&=
M_{{\rm c},0}
+
\int_{\tau_0}^{\tau}
\bar n_{\rm c}(\tau')\,d\tau',
\\
\omega_{\rm c}(\tau)
&=
\omega_{{\rm c},0}
+
\int_{\tau_0}^{\tau}
\dot\omega_{\rm c}(\tau')\,d\tau',
\\
\Omega_{\rm c}(\tau)
&=
\Omega_{{\rm c},0}
+
\int_{\tau_0}^{\tau}
\dot\Omega_{\rm c}(\tau')\,d\tau',
\\
\iota_{\rm c}(\tau)
&=
\iota_{{\rm c},0}
+
\int_{\tau_0}^{\tau}
\dot\iota_{\rm c}(\tau')\,d\tau'.
\label{eq:carrier_flow_angles_explicit}
\end{align}
\subsubsection{Stationary phase approximation}
We require a strategy to evaluate resonant crossings, which mix phase information from the binary and the environment. Recall the binary carrier phase at resonance:
\begin{align}
\psi_l^{(q)}(\tau)
\equiv
-\sigma_l^{(q)}(\tau)\tau
+
lM_{\rm c}(\tau)
+
q\omega_{\rm c}(\tau),
\label{eq:rr_psi_def}
\end{align}
which is a slowly varying quantity within an adiabatic window. For a generic resonant mode $\chi_{\rm p}$, we separate the amplitude and phase of the environmental spectrum. Schematically:
\begin{align}
\langle
\dot\chi_{\rm p}
\rangle_{\mathcal I}^{\rm res}
\propto\exp\!\left[
i\psi_{\rm env}(\tau)+i\psi_l^{(q)}(\tau)
\right],
\label{eq:chi_force_spectrum_def}
\end{align}
where the environmental phase $\psi_{\rm env}$is the argument of the appropriate force spectral coefficient at the resonant frequency. Efficient secular growth occurs when this total phase is locally stationary. For each mode, we define the frequency:
\begin{align}
\Gamma_{l}^{(q)}(\tau)
\equiv
\frac{d}{d\tau} \psi_{\rm env}(\tau)+\frac{d}{d\tau}\psi_l^{(q)}(\tau),
\end{align}
Then, for each mode the stationary points $\tau_{\rm sp}$ satisfy:
\begin{align}
\Gamma_{l}^{(q)}(\tau_{\rm sp})
=
0,
\end{align}
with coherence time:
\begin{align}
t_{l}^{\rm coh}(\tau_{\rm sp})
\equiv
\left|
\frac{d}{d\tau}
\Gamma_{l}^{(q)}(\tau_{\rm sp})
\right|^{-1/2}.
\label{eq:chi_coherence_time_mode}
\end{align}
Assuming isolated and non-degenerate stationary points, the integrated response is given by the sum over all stationary points, with amplitudes given by the stationary phase approximation:
\begin{align}
\int
\langle
\dot\chi_{\rm p}
\rangle_{\mathcal I}^{\rm res}
(\tau')
\,d\tau'
&\propto
\sum_{\rm sp}
\sqrt{2\pi}\,
\mathcal H_{l}^{(q)}\,
t_{l}^{\rm coh}
\exp\!\left[
i\psi_{\rm env} + i\psi_l^{(q)}
+
i\frac{\pi}{4}
\mathrm{sgn}\!\left(
\frac{d\Gamma_{l}^{(q)}}{d\tau}
\right)
\right]\bigg\rvert_{\tau=\tau_{\rm sp}},
\label{eq:spa_integrated_response}
\end{align}
where $\mathcal H_l^{(q)}$ are the appropriate Hansen weighted force spectral amplitudes and all terms are evaluated at the stationary points. 

\subsubsection{Renormalised carrier orbit}

Over the radiation-reaction timescale $T_{\rm rr}$, environmental perturbations to the orbital elements alter the instantaneous emission of gravitational waves, creating a coupled back-reaction. In addition, the secular perturbations to both phases and structural elements can accumulate to order unity, so a strict linear expansion around an unperturbed vacuum track is no longer globally accurate. Both of these issues can be resolved by integrating the equations with the full perturbed orbit on the right hand side. We therefore use the elements without subscripts to denote a ``renormalised'' physical orbit:
\begin{align}
p
&\equiv
p_{\rm c}
+
p_{\rm p},
&
e
&\equiv
e_{\rm c}
+
e_{\rm p},
\\
M
&\equiv
M_{\rm c}
+
M_{\rm p},
&
\omega
&\equiv
\omega_{\rm c}
+
\omega_{\rm p},
\\
\Omega
&\equiv
\Omega_{\rm c}
+
\Omega_{\rm p},
&
\iota
&\equiv
\iota_{\rm c}
+
\iota_{\rm p}.
\label{eq:renormalized_elements_no_superscript}
\end{align}
The evolution equations are then evaluated on these renormalised elements. This absorbs the accumulated perturbations into the carrier itself and yields a closed system of ODEs for the physical inspiral, including both the direct environmental forcing and the induced gravitational-wave back-reaction. 
Note that the same argument must be applied to the perturbing accelerations $(R,S,W)$.

\subsection{ODEs}
\label{Appendix:Resonant_response:ODEs_limits}
\subsubsection{Final ODEs}
Collecting the results above, the secular evolution of the physical binary is governed by the following coupled system:
\begin{align}
\dot p
&=
\dot p_{\rm rr}(p,e)
+
\langle \dot p_{\rm p} \rangle_{\mathcal I}^{\rm res}
\\
\dot e
&=
\dot e_{\rm rr}(p,e)
+
\langle \dot e_{\rm p} \rangle_{\mathcal I}^{\rm res}
\\
\dot\iota
&=
\dot\iota_{\rm PN}(p,e)
+
\langle \dot\iota_{\rm p} \rangle_{\mathcal I}^{\rm res},
\\
\dot\Omega
&=
\dot\Omega_{\rm PN}(p,e)
+
\langle \dot\Omega_{\rm p} \rangle_{\mathcal I}^{\rm res},
\\
\dot\omega
&=
\dot\omega_{\rm PN}(p,e)
+
\langle \dot\omega_{\rm p} \rangle_{\mathcal I}^{\rm res},
\\
\dot M
&=
\bar n(p,e)
+
\langle \dot M_{\rm p} \rangle_{\mathcal I}^{\rm dir}
+
\langle \dot M_{\rm p} \rangle_{\mathcal I}^{\rm geom}.
\label{eq:final_ode_system}
\end{align}
\section{Tables of Hansen Coefficients}
\label{Appendix:Hansen_coeff}

The definition of Hansen coefficients is:
\begin{align}
\left(\frac{r}{a}\right)^j e^{ik\nu}
=
\sum_l X_l^{j,k}(e)e^{ilM},
\end{align}
We also define:
\begin{align}
\mathcal C_l^{j,1}
=
\frac{1}{2}
\left(
X_l^{j,1}+X_l^{j,-1}
\right),
\qquad
\mathcal S_l^{j,1}
=
\frac{1}{2i}
\left(
X_l^{j,1}-X_l^{j,-1}
\right).
\end{align}
Up to cubic order in eccentricity, the coefficients required in Eqs.~\ref{eq:lpe_p_fourier}--\ref{eq:lpe_omega_fourier} are:

\begin{align}
X_l^{1,0}
&=
\begin{cases}
1+\frac{1}{2}e^2+\mathcal O(e^4), & l=0,\\
-\frac{1}{2}e+\frac{3}{16}e^3+\mathcal O(e^5), & l=\pm1,\\
-\frac{1}{4}e^2+\mathcal O(e^4), & l=\pm2,\\
-\frac{3}{16}e^3+\mathcal O(e^5), & l=\pm3,\\
\mathcal O(e^4), & |l|\geq4.
\end{cases}
\end{align}
The coefficients entering the out-of-plane terms are:
\begin{align}
X_l^{1,1}
&=
\begin{cases}
\frac{1}{24}e^3+\mathcal O(e^5), & l=-2,\\
\frac{1}{8}e^2+\mathcal O(e^4), & l=-1,\\
-\frac{3}{2}e+\mathcal O(e^3), & l=0,\\
1-\frac{1}{2}e^2+\mathcal O(e^4), & l=1,\\
\frac{1}{2}e-\frac{3}{8}e^3+\mathcal O(e^5), & l=2,\\
\frac{3}{8}e^2+\mathcal O(e^4), & l=3,\\
\frac{1}{3}e^3+\mathcal O(e^5), & l=4,\\
\mathcal O(e^4), & \text{otherwise},
\end{cases}
\\
X_l^{1,-1}
&=
\begin{cases}
\frac{1}{3}e^3+\mathcal O(e^5), & l=-4,\\
\frac{3}{8}e^2+\mathcal O(e^4), & l=-3,\\
\frac{1}{2}e-\frac{3}{8}e^3+\mathcal O(e^5), & l=-2,\\
1-\frac{1}{2}e^2+\mathcal O(e^4), & l=-1,\\
-\frac{3}{2}e+\mathcal O(e^3), & l=0,\\
\frac{1}{8}e^2+\mathcal O(e^4), & l=1,\\
\frac{1}{24}e^3+\mathcal O(e^5), & l=2,\\
\mathcal O(e^4), & \text{otherwise}.
\end{cases}
\end{align}

\twocolumngrid

\bibliography{hansen}
\bibliographystyle{apsrev4-2}

\end{document}